\documentclass[11pt]{article}
\usepackage[affil-it]{authblk}\usepackage[utf8]{inputenc}
\usepackage{amsfonts} 
\usepackage{amsmath} 
\usepackage{physics} 
\usepackage{color}
\usepackage{float} 
\usepackage{graphicx} 
\usepackage{soul} 

\usepackage{appendix} 
\usepackage{amsthm}   

\newtheorem{rem}{Remark}[section]

\title{Entanglement generation through Markovian feed-back in open two-qubit systems} 
\date{\null}

\author[1,2]{Fabio Benatti}
\author[1,2]{Francesca Gebbia}
\author[3]{Stefano Pisoni}

\affil[1]{Dipartimento di Fisica, Universit\`{a} di Trieste, Strada Costiera 11, I-34151, Trieste, Italy}
\affil[2]{Istituto Nazionale di Fisica Nucleare, Sezione di Trieste, Strada Costiera 11, I-34151, Trieste, Italy}
\affil[3]{Quantum Research Center, Technology Innovation Institute, Abu Dhabi, UAE}

\begin{document}

\maketitle

\abstract{We discuss the generation and the long-time persistence of entanglement in open two-qubit systems whose reduced dissipative dynamics is not apriori engineered but is instead subjected to filtering and Markovian feedback. 
In particular, we analytically study  1.) whether the latter operations may enhance the environment capability of generating entanglement at short times and 2.) whether the generated entaglement survives in the long-time regime.
We show that, in the case of particularly symmetric Gorini-Kossakowski-Sudarshan-Lindblad (GKSL) it is possible to fully control the convex set of stationary states of the two-qubit reduced dynamics, therefore the  asymptotic behaviour of any initial two-qubit state. We  then study 
the impact of a suitable class of feed-back operations on the considered dynamics.}

\section{Introduction}

Quantum entanglement is a fundamental, yet fragile resource in quantum informational tasks~\cite{breuer, nielsen}.
Indeed, it is easily depleted by the presence of noise, typically due to unwanted weak interactions of quantum systems with the environment  within which they are immersed.
If, though weak, such interactions cannot be neglected, these quantum systems are treated as open and are subjected to dissipation and decoherence. In the absence of initial correlations between them and the environment, a reduced dynamics for the open quantum systems alone can be derived through the so-called weak-coupling limit techniques. These latter provide master equations $\partial_t\rho(t)=\mathcal {L}\rho(t)$ of GKSL type~\cite{breuer,Alicki, benatti05, chrushinski17,gorini,lindblad}, where $\rho(t)$ denotes the system state (density matrix)  at time $t\geq 0$.
The generator $\mathcal L=\mathcal{H}+\mathcal{D}$ at the right hand side  gives rise to an irreversible, dissipative Markovian dynamics, 
namely a semigroup of linear maps, formally $\gamma_t:=\exp(t\mathcal L)$, that show no memory effects. The generator 
consists of ($-i$) the commutator $\mathcal{H}$ of the system Hamiltonian $H_S$, 
perturbed by a Lamb-shift $H_{LS}$, with $\rho(t)$, plus a linear dissipative term $\mathcal D$. This latter embodies the noisy and damping effects due to the environment by means of a typical matrix, known as Kossakowski matrix, whose entries are related to the Fourier transforms of the environment two-point time-correlation functions.

If the weak-coupling limit techniques are rigorously applied, on one hand the commutator and the dissipator must commute, $\mathcal H\circ\mathcal D=\mathcal D\circ\mathcal H$, and on the other hand  the Kossakowski matrix results positive semi-definite, which is a necessary and sufficient condition for the generated time-evolution $\gamma_t$ to be completely positive.

This latter property guarantees that the reduced open dynamics, described by  $\gamma_t$, remains physically tenable when lifted to the dynamics $\gamma_t\otimes{\rm id}$ of the open quantum system statistically coupled to a dynamically inert ancilla of any possible finite dimension.
Concretely, the complete positivity of $\gamma_t$, and thus the positive semi-definiteness of the Kossakowski matrix in the generator $\mathcal L$, are necessary and sufficient conditions for the entangled states of the open quantum system coupled to a generic finite-level ancilla to remain \textit{bona fide} quantum states under the factorized joint time-evolution $\gamma_t\otimes{\rm id}$. Namely, for their spectrum to remain positive at all (positive) times and thus for their eigenvalues to keep their interpretation as probability amplitudes.

Typically, the dissipative reduced dynamics of bipartite open quantum systems tends, sooner or later, to destroy any amount of initial entanglement. Indeed, initially separable states are generically sent to separable asymptotic states.
However, it was shown that, by an \textit{ad hoc} suitable engineering of the coupling of two qubits to the environment, one could entangle initially 
separable states~\cite{benatti05,benatti04,benatti05.1,benatti06}. This most desirable effect can be achieved  by purely dissipative means, namely without any environment induced direct dynamical coupling among the single constituents of the bipartite system due to the Lamb-shift Hamiltonian. Moreover, the generated entanglement can be made survive the long-time limit~\cite{benatti05,benatti06}. 

In the following, we assume a different stance with respect to engineering the coupling of two qubits to their environment; namely, we study whether and how gathering information  about the time-evolving open quantum system allows one to operate adjustments to the reduced dynamics that are able to generate entanglement at short times that be also robust against dissipation in the long-time regime. 
These protocols are currently being investigated and are known as filtering~\cite{bouten, gardiner_noise, gough99,wiseman10} and feed-back~\cite{wiseman10,wiseman92,wiseman93}; they consist in monitoring the environment, via \textit{e.g.} homodyne detection, based on whose outcomes modifications of the generator $\mathcal L$ are operated. 

In practice, instead of engineering an \textit{ab initio} appropriate generator $\mathcal L$, in this manuscript we 
analytically investigate case-studies in order to provide insights into the structure of Markovian feed-back protocols  able to dissipatively generate entanglement  when the given reduced dynamics is not.  And, further, to keep it in the long-time 
limit in those cases where, without external intervention,  it would rather disappear. 

Since, in the case of Markovian feedbacks as those studied in what follows, the final effect is to modify the generator to get a physically convenient Kossakowski matrix, the procedure might appear to be a realization of environment engineering; however, one has to notice that the procedure is based on a given environment which is never altered in the course of time, but only monitored. It is only after the monitoring outcomes and according to them that an external intervention on the reduced dynamics of the open system is operated. Though the chosen case-studies are not the most general ones, which would be impossible to handle analytically in a readable manner, in particular for what concerns the stationary states and their entanglement properties, they nevertheless offer a rich phenomenology of different possibilities and help to shed light on how to obtain entanglement enhancing through feed-back protocols.

The paper is subdivided as follows: in the first part of Section~\ref{Sec:2} we shortly overview the necessary tools underlying the reduced dynamics of two open qubits, while in the second part we briefly touch upon the basics of filtering and feed-back applying them in the concrete case of a Hamiltonian Markovian feed-back without direct two-qubit interactions.

Section~\ref{Sec:STE} consists of various sub-sections; in the first one, we resume what is known about short-time two-qubit entanglement generation; in the second one, we  resort to a particular class of symmetric generators of the dissipative dynamics that allows  studying  entanglement generation at short times under feed-back protocols that preserve that symmetry.
In the third and fourth parts of the section, the Hamiltonian contribution $\mathcal H$ is chosen such that two-qubit states of the so-called $X$ form  remain of this form in the course of time.

Finally, in Section~\ref{Sec:LTE}, the asymptotic fate of the entanglement generated at short times is studied by characterizing the stationary states of the dissipative dynamics with feed-back considered in the previous section, thus extending the results already present in the literature~\cite{zhang17}.

The Conclusions summarize the findings presented in the previous sections putting  them into perspective.

\section{Markovian quantum feed-back for a two-qubit dissipative dynamics}  
\label{Sec:2}

In this section we shortly review the case of a two-qubit system interacting with an environment. Later we analyze the dynamics of the open quantum system when it is also subjected to a suitable Markovian feed-back based on the external monitoring of the environment. As emphasized in the Introduction, our purpose is the study of feed-back protocols that enhance the role of the environment in generating two-qubit entanglement and in ensuring its long time persistence.

\subsection{Two-qubit master equation without feed-back}
\label{sec:ME_nofeed}
We begin our discussion by introducing the general setting of two uncoupled qubits interacting with the same environment.

We consider the following Hamiltonian, 
\begin{equation}
H=H_S\otimes\mathbb{I}_B +\mathbb{I}_S \otimes H_B + H_{int} ,\label{H}
\end{equation}
where $H_S$ is a two-qubit Hamiltonian, $H_B$  the environment Hamiltonian, $\mathbb{I}_{S,B}$ are the identity operators of the system and the environment respectively, while $H_{int}$ is an interaction Hamiltonian of the form
\begin{equation}
\label{Hint}
H_{int}=\sum_{\alpha=0}^3\left(\Big(\sigma_\alpha\otimes\mathbb{I}_2\Big)\otimes\Phi_\alpha +   \Big(\mathbb{I}_2\otimes\sigma_\alpha\Big)\otimes\Psi_\alpha\right)\ .
\end{equation}
In the above expression, $\mathbb{I}_2=\sigma_0$ is the qubit identity operator, $\sigma_j$, $j=1,2,3$, the Pauli matrices, while $\Phi_\alpha$, and $\Psi_\alpha$ are suitable Hermitian environment operators that couple the qubits to the environment, possibly not in the same way.
We assume such a coupling to be weak and we absorb the dimensionless coupling strength $\lambda\ll 1$ into the environment operators, for simplicity. 

Then, considering initial factorized states $\rho\otimes\rho_E$, where $\rho$ is a two-qubit density matrix and $\rho_E$ is an equilibrium state of the environment, the rigorous application of the so-called weak-coupling limit procedure yields a GKSL master equation~\cite{benatti05, benatti04,  benatti05.1}:
\begin{equation}
    \frac{\partial \rho(t)}{\partial t}=\mathcal{H}\rho(t)\, +\,\mathcal{D}\rho(t)\equiv \mathcal{L}\rho(t)\ .
    \label{GKSL}
\end{equation}
The generator $\mathcal{L}$ consists of a Hamiltonian term
\begin{equation}
\label{Hamgen}
    \mathcal{H}\rho(t)=-i\left[H , \rho(t)\right], 
\end{equation}
where $H=H_S+H_{LS}$, with
\begin{equation}
\label{Hameff}
H_{LS}=-\frac{1}{2}\sum_{ij} \left(H^{(11)}_{ij}(\sigma_{i}\sigma_j\otimes \mathbb{I}_2)+H^{(22)}_{ij}(\mathbb{I}_2\otimes \sigma_i\sigma_j)+H^{(12)}_{ij}(\sigma_i\otimes\sigma_j)\right)
\end{equation}
a Lamb-shift environment-induced correction of order $\lambda^2$ that, if some of the coefficients $H^{(12)}_{ij}\neq 0$, may induce a dynamical coupling of the two qubits even if $H_S$ does not 
contain any interaction term between them.
Instead, $\mathcal{D}$ in~\eqref{GKSL} is a purely dissipative contribution, also of order $\lambda^2$ to the generator $\mathcal{L}$ that mixes the degrees of freedom of the qubits, 
\begin{equation}
\mathcal{D}\rho=\sum_{\alpha, \beta=1}^{6} K_{\alpha \beta}\left[F_{\beta}\,\rho\,F_{\alpha}-\frac{1}{2}\left\{F_\alpha F_{\beta}, \rho\right\}\right],
\label{D_benatti}
\end{equation}
where $\{\cdot\, , \cdot\}$ denotes the anti-commutator and
\begin{equation}
\label{Fops}
F_{\alpha}:=\sigma_{\alpha} \otimes \mathbb{I}_2\ ,\quad\alpha=1,2,3\ ;\qquad
F_{\alpha}:=\mathbb{I}_2 \otimes \sigma_{\alpha-3}\ ,\quad\alpha =4,5,6\ .
\end{equation}
The coefficients $K_{\alpha\beta}$ are the Fourier transforms of the environment two-point equilibrium time-correlation functions ${\rm Tr}\Big(\rho_E\Phi_\alpha\Psi_\beta(t)\Big)$ and constitute a $6\times6$ positive-semi-definite (Kossakowski) matrix $K=[K_{\alpha\beta}]\geq 0$~\cite{Alicki}. This latter property ensures the full physical consistency of the generated semigroup, namely the so-called complete-positivity of the maps $\displaystyle\gamma_t={\rm e}^{t\mathcal{L}}$.
The Kossakowski matrix can be conveniently put in the form 
\begin{equation}
K=\left(\begin{array}{cc}
\mathcal{A} & \mathcal{B} \\
\mathcal{B}^{\dagger} & \mathcal{C}
\end{array}\right)\geq 0\ ,
\label{K}
\end{equation}
by means of the $3\times 3$ matrices $\mathcal{A}=\mathcal{A}^{\dagger}\geq 0$, 
$\mathcal{C}=\mathcal{C}^{\dagger}\geq 0$ and $\mathcal{B}$.
If $\mathcal{B}\neq 0$, it gives rise to a mixing of the different qubit degrees of freedom that may thus be responsible for the dissipative generation of entanglement. 
Indeed, when $\mathcal{B}=0$ and without couplings between the qubits, the dissipative dynamics would factorize: $\displaystyle\gamma_t={\rm e}^{t\mathcal{L}}={\rm e}^{t\mathcal{L}_1}\otimes{\rm e}^{t\mathcal{L}_2}$, with single qubit generators $\mathcal{L}_1$, $\mathcal{L}_2$.

\begin{rem}
\label{rem:HD}
Besides the complete positivity of $\gamma_t$, another consequence of the rigorous application of the weak-coupling limit~\cite{Spohn} is that
the Hamiltonian contribution~\eqref{Hamgen} and the dissipative one~\eqref{D_benatti} to the generator $\mathcal{L}$ must commute:
\begin{equation}
\label{commgen}
\mathcal{H}\circ\mathcal{D}=\mathcal{D}\circ\mathcal{H}\ .
\end{equation}
\end{rem}

By diagonalizing $K=[K_{\alpha \beta}]=U D U^\dag$ in~\eqref{D_benatti}, with a positive semi-definite diagonal matrix $D={\rm diag}\lbrace \lambda_1,\cdots, \lambda_6 \rbrace$, $\lambda_j\geq 0$, and $U$ unitary such that $K_{\alpha\beta}=\sum_{\mu =1}^6 \lambda_\mu U_{ \alpha \mu } U_{\beta \mu}^*$, one gets a diagonal expression for the dissipative part of the generator:
\begin{equation}
\mathcal{D}\rho=\sum_{\mu=1}^6 \mathcal{D}\left[L_\mu\right]\rho\ ,\quad
\mathcal{D}\left[L_\mu\right]\rho:= L_\mu\, \rho L_\mu^\dagger - \frac{1}{2}\lbrace L_\mu^\dagger L_\mu , \rho\rbrace \label{DLrho}
\end{equation}
with so-called Lindblad operators $L_\mu$ given by
\begin{equation}
\label{DDL}
L_\mu:=\sqrt{\lambda_\mu} \,\sum_{\alpha=1}^6 U_{ \alpha\mu}\,F_\alpha\ .
\end{equation}

To summarize, we briefly analyzed the physics behind the structure of a two-qubit Markovian master equation obtained by means of the weak-coupling limit, pointing out which of its terms may be responsible for the statistical coupling of the two qubits, even in the absence of a direct dynamical interaction between them. In the following we show how the master equation changes when we add a measurement-based quantum feed-back protocol.

\subsection{Two-qubit master equation with feed-back}

In the following, the open two qubit dynamics described in the previous section will be further subjected to a quantum control protocol consisting of two steps; the first one, known as quantum filtering, has the environment constantly monitored in order to harvest information about the qubits; the second one, known as quantum feed-back, uses the information gathered from the environment monitoring in order to externally drive the system dynamics. 

Concretely the monitoring plus feed-back procedure can be summarized as follows: for the sake of simplicity,  consider a generator $\mathcal{L}$  with a single Lindblad operator $L_1:=L$ in~\eqref{DLrho}; then, one knows from Quantum It\^o calculus~\cite{parthasarathy92} that such a dissipative generator can be obtained from coupling the two qubits to an environment consisting of a specific Bosonic bath that induces a quantum Brownian 
dynamics on them.
Then, one constantly monitors the Bosonic environment through homodyne detection. This action introduces a  further noise term in the environment induced master equation that takes into account the conditioning of the state of the two qubits upon the measurement outcomes~\cite{bouten}. Let $\rho_c(t)$ denote the conditioned 
two-qubit state. 

Concretely, the output of the measurement is the so-called homodyne photocurrent $I(t)$, which is a real-valued stochastic variable such that~\cite{wiseman02}
\begin{equation}
    I(t)\,dt={\rm Tr}\left[(L+L^\dagger)\rho_c (t)\right]\,dt + dW(t), 
    \label{photoccurent}
\end{equation}
where $L$ is the Lindblad operator characterizing the disspative behaviour and $dW(t)$ is the Wiener stochastic increment with expectation $E[dW(t)]=0$ and variance $E[(dW(t))^2]=dt$.

Introducing the white noise $\xi(t)=dW(t)/dt$, the master equation for the conditioned state of the system subjected to continuous monitoring via homodyne detection takes the form
\begin{align}
&\dot\rho_c(t) =-i\comm{H}{\rho_c(t)}+\mathcal{D}\left[L \right] \rho_c(t) + \sqrt{\eta}\,\xi(t)\,\mathcal{S}\left[ L \right] \rho_c(t)\label{filtering}
\end{align}
with $\mathcal{D}\left[L \right] \rho_c(t)$ as in \eqref{DLrho} and a non-linear term
\begin{align}
    \mathcal{S}\left[ L \right]\rho_c(t):= L\rho_c(t) + \rho_c(t) L^\dagger - {\rm Tr}\left[( L +L^\dagger )\rho_c(t) \right] \rho_c(t) .
\end{align}
The parameter $0\leq\eta\leq 1$ takes into account the efficiency of the measurement process: in the ideal case $\eta=1$, while $\eta=0$  retrieves an ordinary master equation without detection.

The monitoring of the environment which leads to the previous stochastic master equation is then followed  
by a feed-back protocol; it consists in adding to the generator on the right hand side of \eqref{filtering} a further contribution, $\left[\dot \rho_{c}(t)\right]_{f}$, given by an arbitrary  super-operator $\mathcal{F}$ that may generically depend on all the previous homodyne outcomes~\cite{wiseman10}:
\begin{equation}
\left[\dot \rho_{c}(t)\right]_{f}\equiv\mathcal{F}\left[t, I_{\left[0,t\right[} \right]  \rho_{c}(t)\, , \label{feed-back_mostgeneral}
\end{equation}
where $ I_{\left[0,t\right[}=\lbrace I(\tau)\, | \,\tau \in \left[ 0, t \right[ \rbrace$ is the collection of all the previous measurements.

\begin{rem}
\label{rem:Mark}
Among the different types of feed-back protocols, the most common ones are the so-called Bayesian 
and Markovian feed-backs~\cite{wiseman10,wiseman02}.
The Bayesian feed-back is time non-local in the sense that it drives the open quantum system according to the whole information gathered during a certain interval of time and  takes into account the delay between the measurement process and the consequent feed-back action. It is called Bayesian because the state of the subsystem is constantly updated as soon as more information are gathered. Instead, in the Markovian feed-back, the driving is based only upon the last measurement outcome. 
\end{rem}

In the following we shall adopt instantaneous Markovian feed-backs, with no delay with respect to data acquisition from monitoring the environment; furthermore, the feed-back will be implemented in Hamiltonian form. Namely, the correction to be added to the right-hand side of~\eqref{filtering} will be of the form
\begin{equation}
\label{Mfeed-back}
\left[\dot \rho_{c}(t)\right]_{f}=I(t)\mathcal{K}\left[\rho_c(t)\right]\, , \qquad \mathcal{K}\left[\rho\right]=-i\comm{F}{\rho},
\end{equation}
where $F=F^\dag$ is an arbitrary Hermitian two-qubit operator.  

Using stochastic calculus techniques \cite{wiseman92}, one recovers a master equation where the conditional state $\rho_c(t)$ is statistically independent from the white noise $\xi(t)$. 
Then, averaging over the white noise finally yields the following linear master equation for the density matrix $\rho(t):=E\left[\rho_c(t)\right]$: 
\begin{align}
\label{FeedME}
\dot{\rho}(t)=- i \left[ H + H_{FB}, \, \rho(t) \right]+ \mathcal{D} \left[ L-iF \right] \rho(t) + \frac{1-\eta}{\eta} \mathcal{D} \left[ F \right] \rho(t) ,
\end{align}
where 
\begin{equation}
\label{HFB}
H_{FB}\equiv\frac{1}{2} (L^{ \dagger} F + F L)
\end{equation}
is a feed-back correction to the open system Hamiltonian, while the standard dissipator $\mathcal{D}[L]$ changes to 
$\mathcal{D}[ L-iF]$ (see~\eqref{DLrho}) and, in the non ideal case,  acquires a further dissipative term 
$\mathcal{D} \left[ F \right]$.

\begin{rem}
\label{rem:homod}
Notice that although the measurement process is performed on the environment, yet the interaction of the latter with the qubit system induces a mixing of their degrees of freedom. Indeed, one sees that the actually measured homodyne photo-current operators in \eqref{photoccurent} also depend on the Lindblad operator $L$ of the qubit system. 
In the general case of several Lindblad operators, gathering information about the two qubits through monitoring the environment in the course of time would then immediately be hampered by non-commuting photo-current operators. This problem is avoided by restricting the entire protocol to involve only one, say the first, Lindblad operator. 
\end{rem}

With such a proviso, the generalization of~\eqref{FeedME} is straightforward:
\begin{align}
\nonumber
\dot\rho(t)&= - i \left[ H +\frac{1}{2} (L^{(1) \dagger} F + F L^{(1)}), \, \rho(t) \right] +\mathcal{D} \left[ L^{(1)}-iF \right] \rho(t)\\ 
&+ \frac{2-\eta}{\eta} \mathcal{D} \left[ F \right] \rho(t)\,+\,\sum_{\mu =2}^6 \mathcal{D} \left[ L^{(\mu)}\right] \rho \ . 
\label{feed-back_mostgeneral2}
\end{align}
We know from~\eqref{DDL} and~\eqref{Fops} that the Lindblad operators $L_\mu$ are of the form
\begin{equation}
\label{Krausops}
L^{(\mu)}=\sum_{i=1}^3 \ell_i^{(\mu)}\, \sigma_i \otimes \mathbb{I}_2 + \sum_{i=1}^3 r_i^{(\mu)}\, \mathbb{I}_2 \otimes \sigma_i .\\
\end{equation}
The entries of the matrix $K$ 
in~\eqref{K} are related to $\ell^{(\mu)}_j, r^{(\mu)}_j$ by 
\begin{equation}
\mathcal{A}_{ij}  := \sum_{\mu=1}^6 \ell_i^{(\mu)}\ell_j^{(\mu)*}\ ,\ 
\mathcal{B}_{ij}  := \sum_{\mu=1}^6 \ell_i^{(\mu)}r_j^{(\mu)*}\ ,\ \mathcal{C}_{ij}  := \sum_{\mu=1}^6 r_i^{(\mu)}r_j^{(\mu)*}, \label{ABCtilde}
\end{equation}
and are insensitive to the global phase changes
\begin{equation}
\label{phases}
\ell^{(\mu)}_j\mapsto {\rm e}^{i\varphi_\mu}\, \ell^{(\mu)}_j\ ,\quad
r^{(\mu)}_j\mapsto {\rm e}^{i\varphi_\mu}\,r^{(\mu)}_j\ .
\end{equation}

In order to inspect the entangling properties of the dissipation as a reaction to the feed-back, we choose the latter so that it cannot directly couple the two-qubits; namely:
\begin{equation}
    \label{feed-backop}
F= \sum_{i=1}^3 f_i\, \sigma_i \otimes \mathbb{I}_2 + \sum_{i=1}^3 g_i\, \mathbb{I}_2 \otimes \sigma_i \quad {\rm with} \,\,f_i=f_i^*,\,\,g_i=g_i^* . 
\end{equation}
Then, we obtain a master equation of the same form of \eqref{H}, where the Hamiltonian part is given by the commutator in \eqref{feed-back_mostgeneral2} and the dissipator takes the same form as in~\eqref{D_benatti} with a new Kossakowski matrix $K+\widetilde{K}$, where
\begin{equation}
\widetilde K=\left(\begin{array}{cc}
\mathcal{\widetilde A} & \mathcal{\widetilde B} \\
\mathcal{\widetilde B}^{\dagger} & \mathcal{\widetilde C}
\end{array}\right)\ .
\end{equation}
The feed-back induced entries are explicitly given by
\begin{align}
    &\mathcal{\widetilde A}_{ij}:=  \frac{1}{\eta}\,f_i f_j \,+\, i \ell_i^{(1)} f_j - i f_i \ell_j^{(1)*}, \nonumber\\
\label{feedbachmat}
&\mathcal{\widetilde B}_{ij}:=  \frac{1}{\eta}\,f_i g_j \, +\, i \ell_i^{(1)} g_j - i f_i r_j^{(1)*},\\
    &\mathcal{\widetilde C}_{ij}:=  \frac{1}{\eta}\,g_i g_j \,+\, i r_i^{(1)} g_j - i g_i r_j^{(1)*}. \nonumber
\end{align}
With no feed-back, $f_i=g_i=0$, $i=1,2,3$, certainly $\widetilde{K}=0$; moreover, unlike the original Kossakowski matrix $K$, the feed-back correction $\widetilde{K}$ is sensitive to the phase-changes~\eqref{phases}.
Furthermore, the feed-back correction to the Hamiltoninan contribution explicitly reads
\begin{align}
H_{FB}=\sum_{i,j=1}^3 \left(2\, {\rm Re}\lbrace \ell_i^{(1)} \rbrace\, f_j\, +\, \epsilon_{ijk}\, {\rm Im}\lbrace \ell_i^{(1)}\rbrace\, f_j\,  \Sigma_k\, +\,{\rm Re}\lbrace \ell_i^{(1)} \rbrace\, f_j\, S_{ij}\right)\ .
\label{hamiltonian_fb}
\end{align}
Therefore, a feed-back protocol, besides single qubit corrections, generically introduces a dynamical coupling,
\begin{equation}
\label{FBcoupl}
\widetilde{H}^{(12)}\equiv\sum_{i,j=1}^3 {\rm Re}\lbrace \ell_i^{(1)} \rbrace\, f_j\, S_{ij}\ ,
\end{equation}
of the two qubits that is able to generate two-qubit entanglement and is also sensitive to global phase-changes.

\begin{rem}
\label{rem:FBcorr}
Because of the diagonal Lindblad structure of~\eqref{feed-back_mostgeneral2}, the new Kossakowski matrix $K+\widetilde{K}$ is automatically positive semi-definite. However, despite the fact that the positive pre-factor $1/\eta$ in~\eqref{feedbachmat} can become large in the case of too rough homodyne detections, the correction $\widetilde{K}$ need not in general itself be positive semi-definite. Indeed, with 
$\vert u\rangle=(u_1,u_2,u_3)^T,\vert \ell\rangle=(\ell^{(1)}_1,\ell^{(1)}_2,\ell^{(1)}_3)^T\in\mathbb{C}^3$ and $\vert f\rangle=(f_1,f_2,f_3)^T\in\mathbb{R}^3$, where $^T$ denotes transposition, one gets 
$$
\langle u\vert\widetilde{\mathcal{A}}\vert u\rangle=\frac{1}{\eta}\Big\vert\langle f\vert u\rangle\Big\vert^2+2\,{\mathcal{I}}m\,\Big(\langle f\vert u\rangle\,\langle \ell\vert u\rangle\Big)\ .
$$
Therefore, given any $\eta>0$, by suitably choosing $\vert u\rangle$, with $\vert f\rangle$ of sufficiently small norm, the imaginary part can be made negative and larger in absolute value than the first contribution; hence, $\widetilde{\mathcal{A}}$ is not positive semi-definite.
\end{rem}

Before discussing entanglement generation, we briefly summarize Section 2. We first described the generalities of two-qubit master equations in Section \ref{sec:ME_nofeed} and we proceeded in Section 2.2 explicitly showing how a suitably chosen quantum driving control protocol modifies such master equations. Moreover, we analysed the new terms  arising in the master equation due to the feed-back action, showing which ones may contribute to the entanglement generation between the two qubits.

We continue in Section \ref{Sec:STE} by providing the formalism and the tools to actually check the entanglement formation, applying them to a particular example.

\section{Short-time bipartite entanglement}
\label{Sec:STE}

For two-qubit states $\rho$ evolving according to a dynamics $\gamma_t={\rm e}^{t\mathcal{L}}$, the presence of entanglement is identified by the lack of positivity of the partial transposition of the state $\gamma_t\rho$ and the amount of two-qubit entanglement is quantified by the so-called concurrence \cite{wootters}. The terminology \textit{short-time entanglement generation} stems from the lack of positivity of the partial transposition in the small-time expansion $\gamma_t\rho\simeq\rho\,+\,t\,\mathcal{L}\rho$ or, equivalently, from a positive first order derivative of the concurrence at time $t=0$, for states whose initial concurrence vanishes.

The structure of this Section is as follows: firstly, in Section \ref{sec:ent_condition}, we implement the mathematical conditions for the entanglement generation at short-times, studying how the feed-back action can improve it. 
Secondly, in Section \ref{sec:short-time}, we discuss an analytical instance where the entanglement is enhanced by the feed-back action, considering  the dynamics generated by a symmetric generator.
It turns out that such a generator preserves the structure of a particular set of initial states, the so-called X-states, that we later characterize in Section \ref{sec:X-states}. 
We finally briefly consider the preservation of the entanglement in the long-time regime, which will be the main topic of the last Section of this work.

\subsection{Short-time entanglement generation}
\label{sec:ent_condition}
The aim of this section is to provide the mathematical conditions for entanglement generation at short-times. As a matter of fact, if entanglement is not generated at short-times, it will never be generated. We firstly consider the dynamics generated by~\eqref{GKSL} without any control and feed-back. Then, later in the Section, we show how the entanglement generation conditions change due to the feed-back action.

In order to study the ability of the environment to generate entanglement, one starts with an initial separable pure two-qubit state.

\begin{rem}
\label{rem1}
Notice indeed that, if the environment is not able to entangle any initial state of the form $\rho_S(0)=|\varphi\rangle\langle\varphi|\otimes| \psi\rangle\langle\psi|$, then, because of the semi-group structure of the dynamics, it surely can not entangle separable mixed states at either $t=0$ or at any later time. Indeed, these latter states are, and cannot but evolve into, convex combinations of pure separable states. 
\end{rem}

Given two ortho-normal bases $ \{|\varphi\rangle,|\tilde{\varphi}\rangle\}$ and $\{|\psi\rangle,|\tilde{\psi}\rangle\}$ of the first, respectively second qubit, one can always obtain them by an appropriate rotations of the standard basis $ \{|0\rangle,|1\rangle\}$ of eigenvectors of the Pauli matrix $\sigma_3$:
\begin{align}
&|\varphi\rangle=U|1\rangle, \quad |\tilde{\varphi}\rangle=U|0\rangle , \\
&|\psi\rangle=V|1\rangle, \quad |\tilde{\psi}\rangle=V|0\rangle,
\end{align}
where $U$ and $V$ are unitary operators inducing orthogonal transformations $\mathcal{U}$ and $\mathcal{V}$ of the Pauli matrices:
\begin{align}
U^{\dagger} \sigma_{i} U=\sum_{j=1}^{3} \mathcal{U}_{i j} \sigma_{j}, \quad V^{\dagger} \sigma_{i} V=\sum_{j=1}^{3} \mathcal{V}_{i j} \sigma_{j}.
\end{align} 

Then, as shown in~\cite{benatti05}, the semi-group ${\gamma_t=\rm e}^{t\mathcal{L}}$ generates short-time entanglement in the initial state $|\varphi\rangle\langle\varphi|\otimes| \psi\rangle\langle\psi|$ if and only if
\begin{equation}
\langle u|\mathcal{A}| u\rangle\left\langle v\left|\mathcal{C}^{T}\right| v\right\rangle<\left|\left\langle u\left|\left(\mathcal{R} e(\mathcal{B})+i H^{(12)}\right)\right| v\right\rangle\right|^{2}\label{entanglement_formation},
\end{equation}
where $\displaystyle\mathcal{R}e(\mathcal{B}):=\frac{\mathcal{B}+\mathcal{B}^\dag}{2}$, while $\vert u\rangle$ and $\vert v\rangle$ 
are $3$-dimensional complex vectors with components
 \begin{equation}
 \label{uv}
u_{i}=\sum_{j=1}^{3} \mathcal{U}_{i j}\left\langle 0\left|\sigma_{j}\right|1\right\rangle, \quad v_{i}=\sum_{j=1}^{3} \mathcal{V}_{i j}\left\langle 1\left|\sigma_{j}\right|0\right\rangle.
\end{equation}

As already observed, from \eqref{entanglement_formation} one notices that the entanglement generation depends on the initial state through the vectors $u$ and $v$. Therefore, the $3\times 3$ matrix $\mathcal{B}$ which statistically mixes the two qubits can be responsible for the entanglement generation, as well as the Hamiltonian couplings of the two qubits.

In the presence of Markovian feed-back as in the previous section, the condition~\eqref{entanglement_formation} for initial entanglement generation becomes:
\begin{align}
\langle u|(\mathcal{A}+\mathcal{\widetilde A})|u \rangle\langle v|(\mathcal{C}^T +\mathcal{\widetilde C}^T)|v\rangle &< \big|  \langle u | \mathcal{R}e\lbrace\mathcal{B} + \mathcal{\widetilde B} \rbrace\nonumber\\
&+ i \Big(H^{(12)}+\widetilde{H}^{(12)}\Big) | v \rangle   \big|^2.
\label{entanglement_generation_feed-back}
\end{align}
In the above inequality, the tilde contributions, namely the terms due to the feed-back, are explicitly emphasized.
Notice that, although the feed-back action in~\eqref{feed-backop} has been chosen without two-qubit couplings, it nevertheless induces Hamiltonian couplings of the form~\eqref{FBcoupl} that may contribute to short-time entanglement generation.

To wrap up, entanglement between two non-interacting qubits can be generated by two actions: 1) the Hamiltonian coupling $H^{(12)}$  between the two qubits, which spontaneously arises due to the separate interaction of the qubits with the environment; 2) the correlation functions of the environment embodied in $\mathcal{B}$. However, more interestingly, \eqref{entanglement_generation_feed-back} shows that even in the absence of the terms $H^{(12)}$ and $\mathcal{B}$, entanglement can still be generated by purely dissipative means through a suitable choice of the feed-back parameters hidden in the tilde elements in \eqref{entanglement_generation_feed-back}. Or still, if $H^{(12)}$ and $\mathcal{B}$ do actually produce entanglement, the feed-back protocol can enhance their action.

\begin{rem}
\label{rem:smalleta}
The possibly large contributions to the mixing components $\mathcal{B}+\widetilde{\mathcal{B}}$ of the modified Kossakowski matrix arising from $0\leq\eta\ll 1$ do not contribute to the entanglement generation capability of the environment.
Indeed, using~\eqref{feedbachmat}, when $\eta\simeq 0$, to leading order, the inequality~\eqref{entanglement_generation_feed-back} reads
\begin{equation}
\label{diseq}
\vert\langle u\vert f\rangle\vert^2\,\vert\langle v\vert g\rangle\vert^2\,<\,\left\vert\frac{\langle u\vert f\rangle\langle g\vert v\rangle+\langle u\vert g\rangle\langle f\vert v\rangle}{2}\right\vert^2\ ,
\end{equation}
where $\vert u\rangle$ and $\vert v\rangle$ are vectors in $\mathbb{C}^3$ with components as in~\eqref{uv}, while $\vert f\rangle$ and $\vert g\rangle$ are vectors in $\mathbb{R}^3$ with components $f_i$ and $g_i$ given in~\eqref{feed-backop}.
One can then rewrite
$$
\langle f\vert u\rangle=\tilde{f}_1-i\tilde{f}_2\ ,\quad \langle g\vert v\rangle=\tilde{g}_1-i\tilde{g}_2\ ,
$$
where $\tilde{f}_j=\sum_{i=1}^3\mathcal{U}_{ij}f_i$ and $\tilde{g}_j=\sum_{i=1}^3\mathcal{V}_{ij}f_i$ are generic real numbers for the matrices $[\mathcal{U}_{ij}]$ and $[\mathcal{V}_{ij}]$ are generic orthogonal matrices.
Then,~\eqref{diseq} cannot be satisfied as it amounts to $
\Big(\tilde{f}_1\tilde{g}_2\,-\,\tilde{f}_2\tilde{g}_1\Big)^2\,<\,0$.
\end{rem}

In this Section, we over viewed general conditions for entanglement generation at short-times \eqref{entanglement_formation}. These conditions depend on the initial state of the two-qubit system and on the correlation functions of the environment. We discussed how they change in the presence  of a Markovian feed-back, obtaining \eqref{entanglement_generation_feed-back}, from which it clearly appears how quantum control protocols may alter the capability of the environment to generate entanglement. In the next two sections we give an example of this action considering a more specialized dynamics.

\subsection{Symmetric two-qubit generators}
\label{sec:short-time}
In order to concretely apply the Markovian feed-back protocol previously outlined and to analytically investigate the advantages that can be gained in relation to two-qubit dissipative entanglement generation, we shall consider the case where the coupling to the environment is via the same set of environment operators $\Phi_\alpha=\Psi_\alpha$ in~\eqref{Hint}. Such a choice means that the two qubits experience the presence of the bath in the same way; however, though simplifying, the assumption  nevertheless shows entanglement generation and allows for a full analytical tractability as shown in the case of the Unruh effect in~\cite{benatti04}. Then, the Kossakowski matrix in \eqref{D_benatti} becomes
\begin{align}
    K=\begin{pmatrix}
    \mathcal{A} & \mathcal{A}\\
    \mathcal{A} & \mathcal{A}
    \end{pmatrix},\label{AAAA}
\end{align}
with four 3$\times$3 identical blocks $\mathcal{A}=[\mathcal{A}_{ij}]$. This symmetric form arises because now all the entries of the Kossakowski matrix are the Fourier transforms of ${\rm Tr}\Big(\rho_E\Phi_\alpha\Phi_\beta(t)\Big)$.  

Consequently, the two-qubit dissipative dynamics without feed-back is generated by the following master equation: 
\begin{equation}
\mathcal{L}\rho(t)=-i \left[H,\rho(t)\right]+\sum_{i, j=1}^{3} \mathcal{A}_{i j}\left[\Sigma_{j} \rho(t) \Sigma_{i}-\frac{1}{2}\left\{\Sigma_{i} \Sigma_{j}, \rho(t)\right\}\right]\label{ME_benatti_asy}
\end{equation}
where we have introduced the symmetric single qubit operators 
\begin{align}
&\Sigma_{i}:=\sigma_{i} \otimes \mathbb{I}_2+\mathbb{I}_2 \otimes \sigma_{i}, \quad i=1,2,3\label{Sigma}\ .
\end{align}
These, together with the symmetric two-qubit operators 
\begin{align}
&S_{i j}:=\sigma_{i} \otimes \sigma_{j}+\sigma_{j} \otimes \sigma_{i}, \quad i, j=1,2,3,\label{Sigma_S}
\end{align}
and their anti-symmetric counterparts, provide a linearly independent set of Hilbert-Schmidt orthogonal $4\times 4$ matrices that span the $4\times4$ matrix algebra $M_4(\mathbb{C})$ over $\mathbb{C}^4$.

Since the Kossakowski matrix $K$ is positive semi-definite, such must also be the $3\times 3$ matrix $\mathcal{A}$. The latter can be decomposed into the sum of its symmetric and anti-symmetric parts, 
\begin{equation}
    \label{symdec}
\mathcal{A}=A+B\ ,\quad A=\frac{\mathcal{A}+\mathcal{A}^T}{2}\ ,\quad B=\frac{\mathcal{A}-\mathcal{A}^T}{2}\ ,
\end{equation}
whereby the positivity of $\mathcal{A}$ implies that $A$ is real-symmetric and positive itself. Furthermore, the anti-symmetric component $B$ can always be recast as
\begin{equation}
\mathcal{A}_{i j}=A_{i j}+i \sum_{k=1}^{3} \varepsilon_{i j k} b_{k}\ ,\quad b_k\in\mathbb{R}\ .
\end{equation}

\begin{rem}
\label{rem:diag}
The symmetric matrix $A$ can always be diagonalized by an orthogonal matrix $V=({\bf v_1},{\bf v_2},{\bf v_3})$ with ${\bf v_i}=(v_{i1},v_{i2},v_{i3})\in\mathbb{R}^3$ orthonormal. Then, contributions of the form $\sum_{i,j=1}^3A_{ij}\Sigma_i\rho\Sigma_j$ transform into
$$
\sum_{i,j,\ell=1}^3v_{\ell i}\,v_{\ell j}\,\Sigma_i\,\rho\,\Sigma_j=\sum_{\ell=1}^3 a_\ell\,\widetilde{\Sigma}_\ell\,\rho\,\widetilde{\Sigma}_\ell\ ,
$$
where $a_\ell\geq 0$ are the eigenvalues of $A$ and the symmetric matrices $\widetilde{\Sigma}_\ell$ are constructed with the matrices $\widetilde{\sigma}_\ell:=\sum_{i=1}^3V_{\ell i}\sigma_i$ which still satisfy the Pauli algebra.
This means that it is no restriction to consider Kossakowski matrices where the symmetric component of $\mathcal{A}$ is diagonal. Notice that when an anti-symmetric component is present, the matrix $\mathcal{A}$ can be in general diagonalized only by unitary non-orthogonal matrices $U$, so that the new matrices $\widetilde{\sigma}_\ell=\sum_{i=1}^3U^*_{\ell i}\sigma_i$ would not even be Hermitian in this case.
On the other hand, in the absence of an entangling Hamiltonian, the presence of an anti-symmetric component of $\mathcal{A}$ is necessary for entanglement generation with a Kossakowski matrix as in~\eqref{AAAA}.
Indeed, with all $H^{(12)}_{ij}=0$ in~\eqref{Hameff} and $\mathcal{B}=\mathcal{C}=\mathcal{A}=A$, the condition~\eqref{entanglement_formation} cannot be fulfilled by any choice of initially separable pure states.
In fact, using the Cauchy-Schwartz inequality and the fact that $\mathcal{R}e(\mathcal{A})=A$, one obtains
$$
\langle u|A| u\rangle\left\langle v\left|A\right| v\right\rangle<\left|\left\langle u\left|A\right| v\right\rangle\right|^{2}\leq \langle u|A| u\rangle\left\langle v\left|A\right| v\right\rangle\ .
$$
On the contrary, even when all $H^{(12)}_{ij}=0$, the presence of $B$ guarantees that entanglement is generated for any initial pure separable state $|\varphi\rangle\langle\varphi|\otimes| \phi\rangle\langle\phi|$ that yields $\vert u\rangle=\vert v\rangle$ in~\eqref{entanglement_formation}.
Using that $\mathcal{A}=A+B$ and $\mathcal{A}^T=A-B$, condition~\eqref{entanglement_formation} becomes
$$
\langle u|(A+B)| u\rangle\left\langle u\left|(A-B)\right| u\right\rangle-\left\langle u\left|A\right| u\right\rangle^{2}=-\langle u|B| u\rangle^2<0\ .
$$
\end{rem}

\noindent
According to what precedes, let $\mathcal{A}={\rm diag}(a_1,a_2,a_3)$ and $b_{1,2,3}=0$ 
so that the dissipative term $\mathcal{D}$ in the generator~\eqref{GKSL}
cannot generate entanglement, while we allow for possible entangling terms in the Hamiltonian. Also, because of~\eqref{ABCtilde}, one can choose $\ell^{(1)}_i=\sqrt{a_i}$ and $\ell^{(\mu)}_i=0$ for $\mu\neq 1$.

Furthermore, as stated in Remark \ref{rem:HD}, the  Hamiltonians $H$ arising from  the weak coupling limit must be such that the corresponding contributions to the generators, in absence of feed-back, commute with the dissipative ones.
As shown in Appendix~\ref{AppB}, this can be achieved  only if the diagonal matrix $A$ is a multiple of the identity, namely only if $a_{1}=a_{2}=a_{3}=a$, and the Hamiltonian has the form
\begin{align}
\label{physH}
H=\alpha \Sigma_{1}+\beta \Sigma_{2}+\gamma \Sigma_{3}+\delta\, S\ ,\qquad S\equiv S_{11}+S_{22}+S_{33}\ ,
\end{align}
where $\alpha,\beta,\gamma,\delta$ are real parameters. 

At this point, we modify the non-entangling dynamics by means of a feed-back protocol that, as explained before, amounts to substituting the Kossakowski matrix $K$ with a new  one, $K+\widetilde{K}$, where the tilde elements are  as in \eqref{ABCtilde}. In order to keep the symmetric structure~\eqref{AAAA}, we choose $f_i=g_i$ in~\eqref{feedbachmat}, namely we operate a feed-back that acts in the same way on both qubits. 
Further, we simplify the feed-back by choosing $f_1=f_3=0$, while leaving $f\equiv f_2$ as a free control parameter.

Altogether, these conditions yield the following dissipator:
\begin{align}
&\mathcal{D}_F\rho(t) = \sum_{i,j=1}^3 (\mathcal{A}_{ij}+\mathcal{\widetilde A}_{ij}) \,\left[ \Sigma_j\,\rho(t)\,\Sigma_i - \frac{1}{2}\Big\lbrace \Sigma_i \Sigma_j,\,\rho(t) \Big\rbrace \right]\quad{\rm with}\nonumber\\
\label{AA}
&\mathcal{A}=[\mathcal{A}_{ij}]=a \mathbb{I}_3, \qquad
\widetilde{\mathcal{A}}=[\mathcal{\widetilde A}_{ij}]=\begin{pmatrix}
0 & i \sqrt{a}\,f & 0 \\
-i \sqrt{a}\,f & f^2 & 0 \\
0 & 0 & 0
\end{pmatrix}.
\end{align}
Under the same conditions, the feed-back correction~\eqref{hamiltonian_fb} to the Hamiltonian contribution explicitly reads 
\begin{equation}
H_{FB} =\,f\,\sqrt{a}\,\Big(2\,+\,S_{12}\Big)\ .
\label{hamiltonian_fb_exp}
\end{equation}

One thus sees that the dissipative contribution to the generator presents an entangling anti-symmetric part in its Kossakowski matrix as well as a two-qubit 
coupling $S_{12}$ in the Hamiltonian part.
Also, in connection with Remark~\ref{rem:FBcorr}, notice that even if $\widetilde{\mathcal{A}}$ is not positive semi-definite, the full Kossakowski matrix is positive. Indeed 
$$
\mathcal{A}+\widetilde{\mathcal{A}}=\begin{pmatrix}
a&i\sqrt{a}f&0\cr
-i\sqrt{a}f&a+f^2&0\cr
0&0&a\end{pmatrix}\geq 0\ .
$$

In this section, we considered a dissipative dynamics unable to generate entanglement, due to lack of off-diagonals terms in the Kossakowski matrix and of two-qubit interaction terms in the Hamiltonian evolution. We then devised a simple feed-back protocol controlled by just one free parameter $f$ and obtained the master equation in \eqref{AA}, where now off-diagonals terms in the Kossakowski matrix are present and clearly depend on $f$, together with interaction terms~\eqref{hamiltonian_fb_exp} in the Lamb-shift corrections to the Hamiltonian, also arising from the feed-back action. Since entanglement generation clearly depends on the initial state of the two-qubit system, in the following Section we restrict our considerations to a particular set of initial states, namely the $X$-states, and to a generator which preserves their structure.

\subsection{$X$-states}
\label{sec:X-states}

In order to be able to analytically discuss various possible scenarios, we shall restrict to Hamiltonians~\eqref{physH} of the form 
\begin{equation}
H=\gamma \Sigma_{3}+\delta S \label{H_gamma_delta} \ .
\end{equation}
The feed-back driven master equation then reads
\begin{align}
\nonumber\frac{\partial \rho(t)}{\partial t}&=-i\left[\gamma \Sigma_{3}+\delta S+\sqrt{a}\,f\, S_{12}, \rho(t)\right]\\
\label{XdissME}
&+\sum_{i, j=1}^{3}\left(\mathcal{A}_{i j}+\widetilde{\mathcal{A}}_{i j}\right)\left[\Sigma_{j} \rho(t) \Sigma_{i}-\frac{1}{2}\left\{\Sigma_{i} \Sigma_{j}, \rho(t)\right\}\right],
\end{align}
with $\mathcal{A}$ and $\widetilde{\mathcal{A}}$ as in~\eqref{AA}.

The reason for setting $\alpha=\beta=0$ in~\eqref{physH} is because then
the master equation~\eqref{XdissME} maps the class of the so-called $X$-states into itself making such states particularly suited for analytical 
considerations~\cite{quesada}.
The $X$-states are two-qubit states that, in the computational basis, take the form:
\begin{align}
\label{X states matrix}
& \rho_X = 
    \begin{pmatrix}
    a & 0 & 0 & w \\
    0 & b & z & 0 \\
    0 & z^* & c & 0 \\
    w^* & 0 & 0 & d
    \end{pmatrix}\ ,
\end{align}
where normalization and positivity of $\rho_X$ ask for
\begin{equation}
\label{constraints on X states}
a+b+c+d=1\ ,\quad a,b,c,d \geq 0\ ,\quad
|z| \leq \sqrt{b c}\ ,\quad |w| \leq \sqrt{a d}.
\end{equation}
Notice that the separable pure states (in the computational basis)
\begin{align}
\rho_1(0)&=|0\rangle\langle 0|\otimes|0\rangle\langle 0|\ ,\qquad
\rho_2(0)=|1\rangle\langle 1|\otimes|0\rangle\langle 0|\ ,\nonumber\\
\rho_3(0)&=|0\rangle\langle 0|\otimes|1\rangle\langle 1|\ ,\qquad
\rho_4(0)=|1\rangle\langle 1|\otimes|1\rangle\langle 1|\ ,
\label{initial_states}
\end{align}
 are $X$-states; while the Bell states,
\begin{align}
\label{BellStates}
\begin{split}
\vert\psi_1\rangle&=\frac{\vert 00\rangle+\vert11\rangle}{\sqrt{2}}\ ,\qquad
\vert\psi_2\rangle=\frac{\vert 00\rangle-\vert11\rangle}{\sqrt{2}}\\
\vert\psi_3\rangle&=\frac{\vert 01\rangle+\vert10\rangle}{\sqrt{2}}\ ,\qquad
\vert\psi_4\rangle=\frac{\vert 01\rangle-\vert10\rangle}{\sqrt{2}}\ ,
\end{split}
\end{align}
are a set of pure and entangled $X$-states.

\begin{rem}
\label{rem:Xstates}
In the so called Fano representation of two qubit states,
\begin{align}
\rho=\frac{1}{4} \left( \mathbb{I}_2\otimes\mathbb{I}_2 + \sum_{i=1}^3 \rho_{0i} \,\mathbb{I}_2\otimes \sigma_i +\sum_{i=1}^3 \rho_{i0} \,\sigma_i \otimes \mathbb{I}_2 +\sum_{i,j=1}^3 \rho_{ij} \,\sigma_i \otimes \sigma_j   \right) \, , \label{fano0} 
\end{align}
$X$-states  involve  the following subset of operators only:
\begin{equation*}
    \mathcal{S}:=\{\mathbb{I}_2 \otimes \mathbb{I}_2,\sigma_3 \otimes \mathbb{I}_2,\mathbb{I}_2 \otimes \sigma_3,\sigma_1 \otimes \sigma_1,\sigma_2 \otimes \sigma_2, \sigma_3 \otimes \sigma_3, \sigma_1 \otimes \sigma_2, \sigma_2 \otimes \sigma_1\} \ ,
\end{equation*}
which is closed under multiplication: $\mathcal{S}\mathcal{S}\mapsto\mathcal{S}$.
The remaining 8 tensor products of Pauli matrices constitute a set
\begin{equation*}
    \mathcal{S}':=\{\sigma_1 \otimes \mathbb{I}_2,\mathbb{I}_2 \otimes \sigma_1,\sigma_2 \otimes \mathbb{I}_2,\mathbb{I}_2 \otimes \sigma_2,\sigma_1 \otimes \sigma_3,\sigma_2 \otimes \sigma_3,\sigma_3 \otimes \sigma_1,\sigma_3 \otimes \sigma_2\}
\end{equation*}
which is such that $\mathcal{S}\mathcal{S}'\mapsto\mathcal{S}'$ while $\mathcal{S}'\mathcal{S}'\mapsto\mathcal{S}$.
Because of this algebraic properties, the generator in~\eqref{XdissME} keeps the form of $X$-states. This is not true for a more general Hamiltonian as in~\eqref{physH}.
Such a property will be useful in the next Sections when we investigate if the entanglement generated at short-times is preserved in the asymptotic limit when the feed-back is switched off (in this case the short-times entanglement generation is only due to the mixing Hamiltonian $S$ term).
\end{rem}

Let us consider the pure and separable states in~\eqref{initial_states} as initial states.
By explicit check of condition~\eqref{entanglement_generation_feed-back} for short-time entanglement generation, one gets
\begin{enumerate}
\item For $\rho_1(0)$ and $\rho_4(0)$:  $a(\sqrt{a}\,-\,f)^2<0$ which is never satisfied.
\item  For $\rho_2(0)$ and $\rho_3(0)$: $-(a\,f^2+4\delta^2 ) <0$ which is always satisfied.\label{2}
\end{enumerate}
Notice that $a\geq 0$ for the Kossakowski matrix must be positive semi-definite.
Hence, there exists a class of initial states, those that are convex combinations of  $\rho_1(0)$ and $\rho_4(0)$, that are never entangled by the dissipative dynamics even when controlled by feed-back, if chosen the way we did. On the other hand, there are states which become entangled thanks to the chosen feed-back and the initial Hamiltonian.

In particular, from the expression in the second item above, we see that there are two contributions to the entanglement generation: one due to the feed-back action, namely $a f^2$, the other one, $4\delta^2$, due to the initial Hamiltonian~\eqref{H_gamma_delta}. Thus, if we set $\delta=0$, in other words if we discard the $S$ operator from \eqref{H_gamma_delta}, yet there is still entanglement generation thanks to the feed-back action.

In line of principle, the feed-back contribution to the entanglement generation, $a f^2$, might be due either to the feed-back correction to the Hamiltonian $\sqrt{a}f\,S_{12}$ in \eqref{hamiltonian_fb_exp} or to the feed-back correction  $\mathcal{\tilde{A}}$  \eqref{AA} to the Kossakowski matrix. However, for the separable initial state we chose, one finds $\langle u | S_{12}|v\rangle=0$, so that the $af^2$ contribution can only be due to the feed-back modified dissipator, confirming the possibility of generating entanglement in a dissipative evolution through the feed-back action.

Here we have focused upon entanglement generation at short times; in the next Section,
we address the question of whether entanglement can persist even asymptotically in time. Before that, we make a short digression about the asymptotic properties of a dissipative dynamics without feed-back.

\subsection{Beyond short-times}
\label{sec:beyond short-time}

Some intuitions about the long-time fate of the initially generated entanglement is gained by studying the evolution generated by (\ref{XdissME}) when the feed-back free parameter $f$ is set equal to zero for a sub-class of X-states and for the Bell states that will serve as benchmark states for the dynamics with feed-back later on.

Notice that, by measuring time in unit of $\gamma$ and setting the dissipative parameter $a=1$, we can always reduce to only one free parameter $\delta$ in~\eqref{XdissME} whose associated two-qubit operator is $S$. Let us then consider one of the initial pure and separable X states in \eqref{initial_states} that gets entangled at short-times, namely
\begin{equation}
\rho_2(0)=|1\rangle\langle 1|\otimes|0\rangle\langle 0| \ .
\end{equation}
When $f=0$, the time evolved state $\rho_2(t)$ is found to be:
\begin{align}
\rho_2(t)=\begin{pmatrix}
A(t)& 0&0&0\\
0&B_+(t)&C_+(t)&0\\
0&C_-(t)&B_-(t)&0\\
0&0&0&A(t)\\
\end{pmatrix},
\end{align}
where
\begin{align*}
A(t)&:=\frac{1}{6}\Big(1-e^{-12 t}\Big)\ ,\quad B_\pm(t):=\frac{1}{6} \Big(\pm 3 e^{-4 t} 
\cos (8 \delta t)+e^{-12 t}+2\Big)\\
C_\pm(t)&:=\frac{1}{6} \Big(-1 + e^{-12 t} \pm 3 i e^{-4 t} \sin(8 \delta t)\Big)\ .
\end{align*}
The amount of entanglement of two-qubit $X$-states as measured by their concurrence~\cite{wootters} can be analytically computed.
In the present case it reads: 
\begin{align}
\nonumber   
\mathcal{C}(\rho_2(t))&=\frac{1}{3} \sqrt{9 e^{-8 t} \sin ^2(8 \delta t)+e^{-24 t}-2 e^{-12 t}+1}\\
   &\hskip 2cm -\frac{1}{3} \sqrt{e^{-24 t} \left(e^{12 t}-1\right)^2} \ .
\end{align}
The following plot shows $\mathcal{C}(\rho_2(t))$ for different values of $\delta$
\begin{figure}[H]
\centering
\includegraphics[scale=0.50]{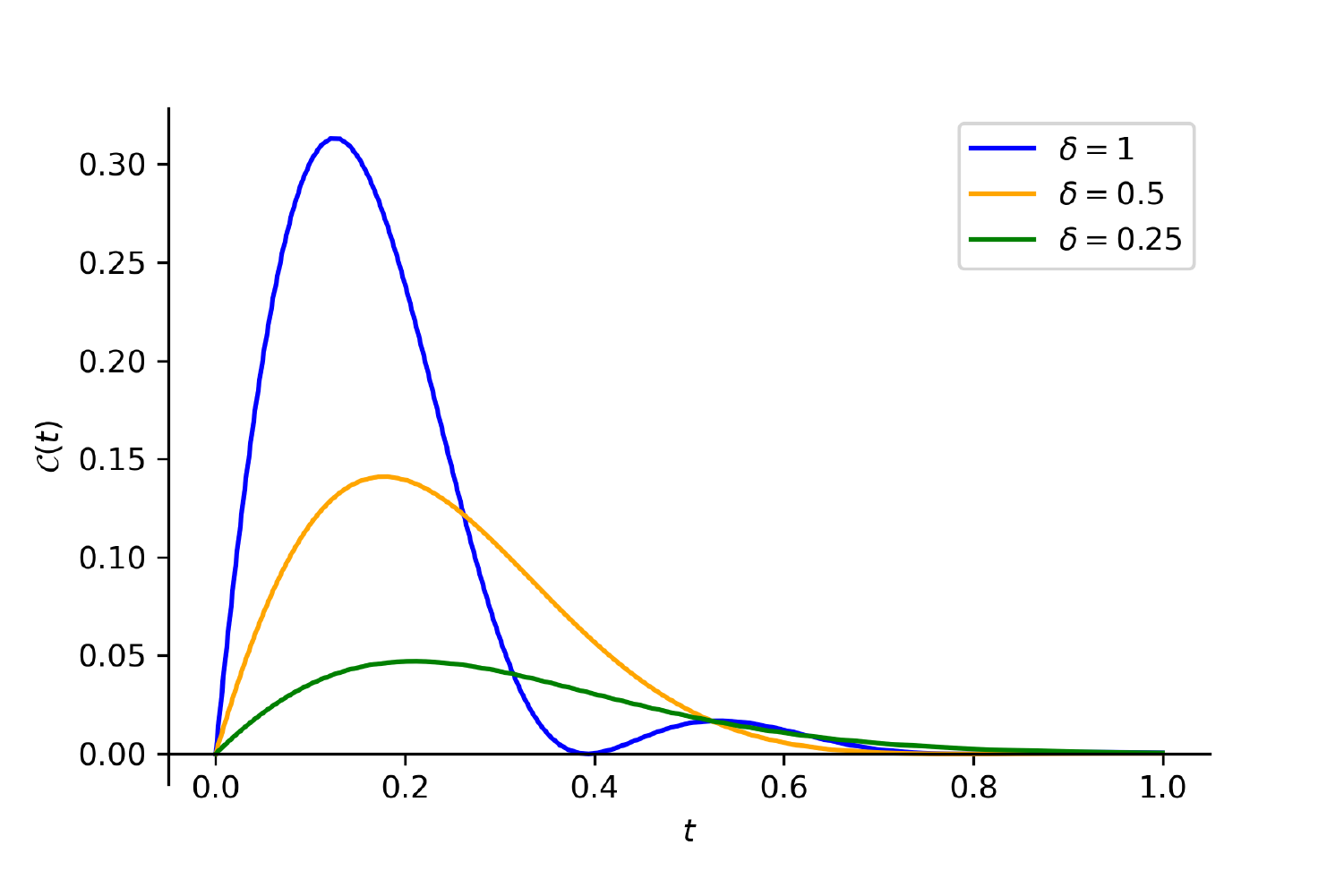}
\caption{\small Concurrence as a function of time.}
\label{FigC}
\end{figure}
\noindent
Entanglement is indeed generated by $S$ at short times, exhibits local minima and maxima which depend on $\delta$ and vanishes asymptotically. A similar concurrence dynamics characterizes the state $\rho_3(0)$.

Analogously, one finds that all Bell states but one are mapped into separable states, the exception being the projector onto $\vert\Psi_4\rangle$ in~\eqref{BellStates}. The latter is indeed  a fixed point of the dynamics and its maximal entanglement remains unaffected by dissipation.

In the following we shall show that, acting with an appropriate feed-back, not only $\rho_2(0)$ and $\rho_3(0)$ get entangled at short-times, but they also remain entangled in the asymptotic regime. 

\section{Two-qubit asymptotic entanglement}
\label{Sec:LTE}
In this final Section, we analytically study the fate of the entanglement generated at short time thanks to the feed-back action on the dissipative dynamics discussed in the previous Section.
In order to do this, the ergodic properties of the semigroup evolution generated by~\eqref{XdissME} need to be analyzed.

On general grounds, the effects of decoherence and dissipation that counteract entanglement production are expected to be dominant at large times, so that no entanglement is left at the end as indeed shown in Fig.\ref{FigC}. However, this need not always be the case as we show in this section. Indeed, there are cases where entanglement never entirely vanishes, yielding entangled stationary states. From now on the asymptotic states will be denoted by $\hat\rho$. They are obtained as solutions to
\begin{equation}
\label{stateq}
\frac{\partial \hat\rho}{\partial t} = \mathcal{L}\hat\rho= 0\ .
\end{equation}
We shall consider generators $\mathcal{L}$ as in the right hand side of~\eqref{XdissME} where  we  set $\gamma=\delta=1$, for sake of simplicity.

It proves useful to work with the Fano representation~\eqref{fano0}; indeed, one can easily check that the quantity $\tau:=\sum_{i=1}^3 \rho_{ii}$, which must satisfy $-3 \leq \tau \leq 1$ for $\rho\geq 0$, is  a constant of the motion. Moreover, in Appendix~\ref{AppA} it is proved that the stationary states of generators involving only qubit-exchange symmetric matrices $\Sigma_j$ and  $S_{ij}$ must be of the form 
\begin{equation} 
\label{symmetric manifold}
\hat\rho = \frac{1}{4} \bigg[ \mathbb{I}_2 \otimes \mathbb{I}_2 + \sum_{i=1}^3 \hat\rho_i \ \Sigma_i + \sum_{i,j=1}^3 \hat\rho_{ij} \ S_{ij} \bigg],
\end{equation}
with $\hat\rho_{ij}=\hat\rho_{ji}$.

Using \cite{frigerio}, the problem of finding the invariant states reduces to seeking an invertible stationary state $\hat\rho_0$ with strictly positive eigenvalues. Given the chosen generator, one such stationary state is 
\begin{align} \label{faithful state}
\hat \rho_0 = \frac{1}{4} \bigg[\mathbb{I}_2 \otimes \mathbb{I}_2 + M\, \Sigma_3 - N\, (S_{11}-S_{22})+R\, S_{33} -L\,S_{12}\bigg]\ .
\end{align}
The explicit dependence of the real coefficients $M,N,R,L$ on the only two remaining parameters, the rate $a$ and  the feed-back parameter $f$, are given in Appendix~\ref{AppA}.

\begin{rem}
\label{rem:stst}
The stationary state in~\eqref{faithful state} is not unique: as we already observed, the Bell projector $P=\vert\Psi_4\rangle\langle\Psi_4\vert$ is $\gamma_t$-invariant, although it is not invertible.
An important result of the theory developed in~\cite{frigerio} is that all initial states $\rho(0)$ tend to a stationary state of the form 
\begin{align} 
\label{manifold}
\hat\rho=\frac{P \hat \rho_0 P}{\mathrm{Tr}[P \hat \rho_0 P]} \mathrm{Tr}[P \rho(0)] + \frac{Q \hat \rho_0 Q}{\mathrm{Tr}[Q \hat \rho_0 Q]}\mathrm{Tr}[Q \rho(0)],
\end{align} 
where $\mathrm{Tr}[P \rho(0)] = \frac{1}{4}(1-\tau)$ and $Q=\mathbb{I}_4-P$.
This follows since the so-called commutant set of the Lindblad operators in $\mathcal{L}$ coincides with the commutant set of the Lindblad operators plus the Hamiltonian and it corresponds to the commutative algebra generated by $\mathbb{I}$ and $S$ in~\eqref{physH}.
The manifold of asymptotic states is then parametrized by the constant of motion $\tau$ and given by the convex combinations of the orthogonal projectors $\displaystyle P=\frac{1}{4} \left( \mathbb{I}_4 - \frac{S}{2} \right)$, $1$-dimensional, and $Q$, $3$-dimensional.
\end{rem}

Plugging \eqref{faithful state} into \eqref{manifold}, the coefficients of the asymptotic states in \eqref{symmetric manifold} are found to be:\\
\begin{align}
    & \hat\rho_3 = \frac{M (\tau +3)}{2 R+3},\quad \hat\rho_{12} = -\frac{L (\tau +3)}{2 R+3}
    \ ,\quad \hat\rho_{11} = \frac{-2 N (\tau +3)-2 R+\tau }{4 R+6},\\
    & \hat\rho_{22} = \frac{+2 N (\tau +3)-2 R+\tau }{4 R+6},\quad
    \hat\rho_{33} = \frac{2 R (\tau +2)+\tau }{4 R+6}\ .
\end{align}
The stationary states depend on the initial condition $\rho(0)$ only through the value of the parameter $\tau$ so that all different initial states with a same $\tau$ tend asymptotically to the same stationary state. This latter is an $X$-state whose concurrence can thus be analytically computed:
\begin{align}
\label{stat-conc}
    \mathcal{C}[\hat\rho] = 2 \ \mathrm{max} \lbrace 0, D_1 , D_2\rbrace,
\end{align}
with $D_{1,2}$ the following functions of $\tau$:
\begin{align}
D_1(\tau)&=  \frac{2\left|\tau-2R\right|-(\tau+3)\sqrt{(1+2R)^2-4M^2}}{4(3+2R)}\ ,\\
D_2(\tau)&= \frac{2(\tau+3)\,\sqrt{4N^2+L^2}-\left|\tau(1+2R)-3+2R\right|}{4(3+2R)}\ .
\end{align}
Plotting $D_2(\tau)$ against $\tau$ shows that it is nowhere positive independently of the parameters $a$ and $f$, 
while $D_1(\tau)$ behaves as shown in Figure~\ref{Figtau}.
\begin{figure}[H]
\centering
\includegraphics[scale=0.28]{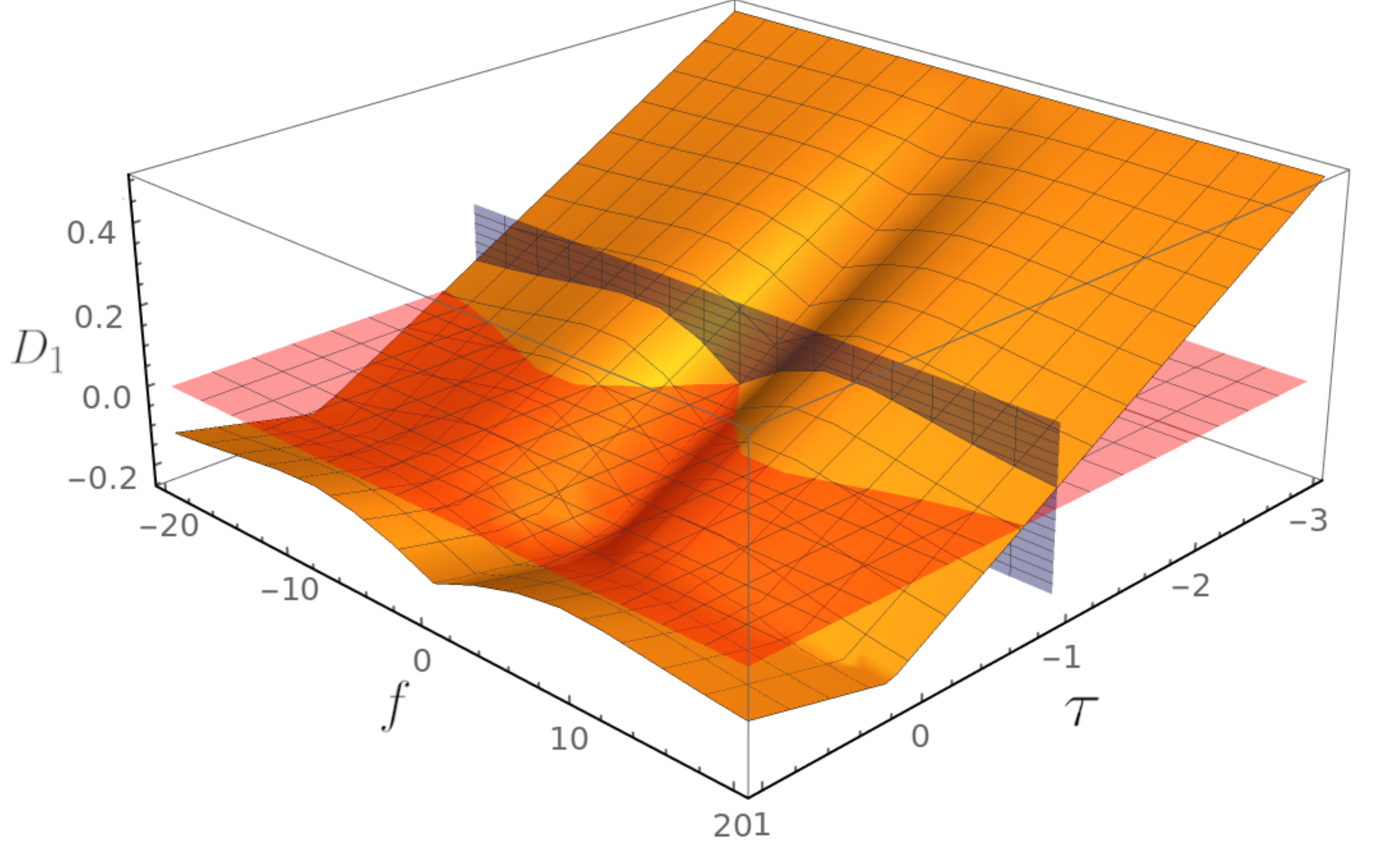}\caption{\small
{$D_1$ against feed-back parameter $f$ and constant of motion $\tau$. 
Red plane: $D_1>0$ becomes bigger than zero (asymptotic entanglement). Blue plane: states with $\tau=-1$.}}
\label{imm_D1}
\end{figure}
In order to unambiguously identify the effects of the feed-back, we compare the case without feed-back ($f=0$) with  the near optimal value\footnote{The optimal value for $f$ lies in the interval $[5,6]$, as one can see from the Fig.~\ref{imm_D1} above.} $f=5$, for fixed value of $a=10$:
\begin{figure}[H]
\centering
\includegraphics[scale=0.28]{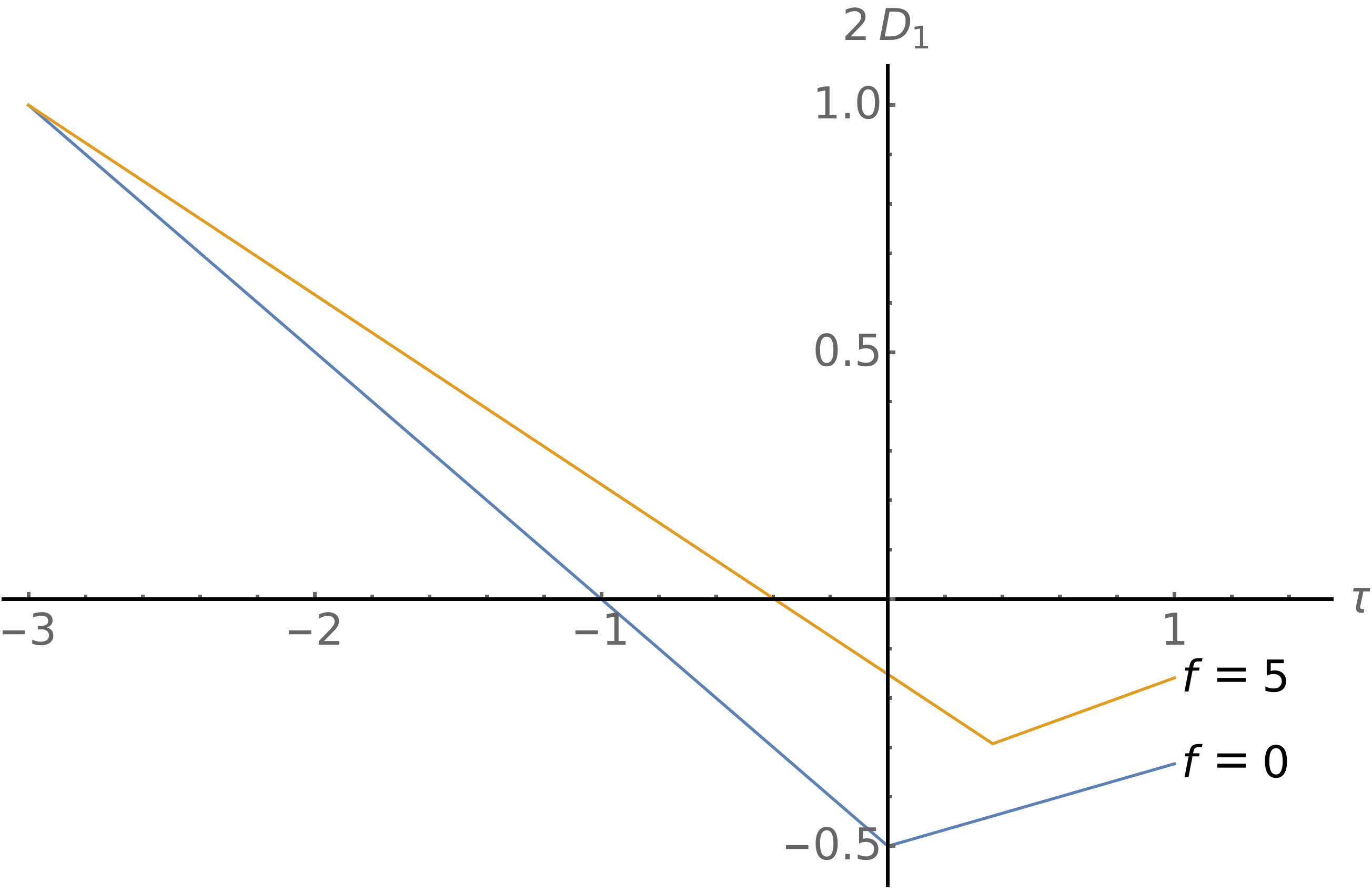}
\caption{\small{$2D_1$ vs $\tau$, for $f=0$, $f=5$ and $a=10$.}}
\label{Figtau}
\end{figure}

From the plot we see that the amount of entanglement in the asymptotic regime can improve for any value of $\tau$ via the feed-back action. Moreover, the value of $\tau$ uniquely determines the amount of entanglement in the stationary states. Then, given any initial state, it suffices to calculate $\tau$ to know if that state will end up to be entangled or not.

In Section \ref{sec:X-states}, the separable states $\rho_2(0)=|1\rangle\langle 1|\otimes|0\rangle\langle 0|$ and $\rho_3(0)=|0\rangle\langle 0|\otimes|1\rangle\langle 1|$ have been shown to become entangled by means of appropriate feed-backs. We now examine if the entanglement generated by the feed-back action at short times survives in the asymptotic regime. Both the states $\rho_2(0)$ and $\rho_3(0) $ happen to have $\tau=-1$. As one can see from from Figure \ref{Figtau}, when $f \neq 0$, the entanglement generated at short times persists also in the long-time regime. Hence, it is the feed-back itself which is responsible for the entanglement generation and its preservation.

It is worth asking what happens to initial states which are already entangled. In the case of the Bell states,  the first three of them in \eqref{BellStates} evolve asymptotically into separable states regardless the feed-back action; indeed, they are characterized by $\tau = 1$ (see Figure \ref{Figtau}). They are thus an instance of initially entangled states that lose their entanglement asymptotically. The same is not true for the fourth Bell state, $ |\psi_4\rangle = (| 01 \rangle - |10 \rangle)/\sqrt{2}$, which needs a separate discussion since on one hand $\tau=-3$ and on the other hand it is a stationary state for the dynamics. Indeed, as shown in Appendix \ref{AppA},
\begin{equation}
        P=\frac{1}{4} \left( \mathbb{I}_4 - \frac{S}{2} \right)=\ket{\psi_4}\bra{\psi_4},
\end{equation}
 is a fixed point of the dynamics that projects onto a separate subspace. Therefore, the maximal entanglement of the fourth Bell state does not get affected and remains maximal as shown in Figure \ref{Figtau} with $\tau = -3$. 

Another feature arises from Figure \ref{Figtau}  when we consider $f=0$ (represented by the blue line). Namely, all  asymptotic states characterized by $\tau < -1 $ are entangled. Since the dynamics is not able to generate entanglement when $f=0$ (as shown in Section \ref{sec:beyond short-time}), the one found in these asymptotic states must be present initially at $t=0$. Since every initial state is mapped into the asymptotic one characterized by the same constant parameter $\tau$, we conclude that all pure bipartite states with $\tau < -1 $ are entangled states. Of course, this is not a necessary condition; for instance, apart from $\vert \psi_4\rangle$, the other three Bell states are entangled and have $\tau=-1$. More in general, $\tau$ alone is not enough to fully characterize the entanglement of a generic bipartite two-qubit state.

\section{Conclusions}
In this work we addressed the issue of entanglement generation and persistence in open quantum
systems that, beside the dissipative effects due to the presence of a suitable environment, also undergo
feed-back actions that are implemented based on the monitoring of the latter.

It was already known that, by suitably engineering the environment, one can entangle initially separable states of bipartite systems immersed in it and make the generated entanglement to persist despite decoherence.
In this work, instead of engineering the coupling of the open system to its environment in order  to obtain a master equation with entangled asymptotic states, we studied the possibility of adjust the dynamics by means of appropriate filtering and feed-back protocols.
In both cases, the final result is a GKSL generator with desired properties, the difference with respect to the first approach is that in the one pursued in this paper 
the environment is not modified but monitored and the generator is changed based on the monitoring outcomes. This change of perspective can take a variety of forms: to start with, we considered Markovian feed-backs. Other possibilities are however available, among which Bayesian feed-backs where memory effects are taken into account. 

In the present paper, we focused upon the dynamics of two open qubits whose interaction with the environment is described  by a master equation of GKSL type that is then altered according to a Markovian feed-back protocol. This protocol makes the dissipative dynamics either able to generate entanglement or to improve the entanglement generation capability of the environment when the master equation already allows it.

In practice, the existing theory about dissipative entanglement generation provides necessary and sufficient conditions for initial separable pure states to become entangled at short times; these conditions depend on the structure of the so-called Kossakowski matrix which characterizes the purely dissipative contribution to the generator of the master equation. Then, in the first part of the manuscript, we applied a specific Markovian feed-back protocol to control a non-entangling open dynamics for two open qubits and showed by means of concrete examples how to obtain a new Kossakowski matrix that can achieve entanglement at short times.

Then, we studied the persistence of entanglement in the long-time regime. In particular, we characterized the convex set of stationary states of a particular class of dissipative dynamics showing how the feed-back parameters can be tuned in order for the entanglement generated at short-times to be preserved asymptotically, when there would be no entanglement without feed-back. 

The results reported thus provide a first series of indications on how to devise and optimize more structured and complex feed-back protocols, \textit{i.e.} with several tunable parameters. Notice that other kinds of feed-back protocols, called Bayesian, may be implemented such that the resulting modified master equations become non-Markovian. Supposedly, such protocols may perform better than the Markovian ones, since they constantly update the open quantum system states depending on the continuous monitoring of the environment. Yet, these feed-back procedures are mostly amenable to numerical studies only and have not been considered in the present manuscript whose purpose was instead to provide as much an analytical insight into the matter as possible.
\vskip .5cm

\noindent
\textbf{Acknowledgements}: FB acknolewdges financial support from PNRR MUR project PE0000023-NQSTI.

\appendix
\appendixpage

\section{Stationary states}
\label{AppA}

In this Appendix, we outline the procedure to construct a faithful state, \textit{i.e.} without null eigenvalues, of the form (\ref{faithful state}) which is left invariant by the dynamics generated by the master equation~\eqref{XdissME} with Kossakowski matrix as in~\eqref{AA}.

The key observation is that the algebra $\{L_\alpha\}'$ consisting of all matrices commuting with the set 
$\{L_\alpha\}$ of Lindblad operators in the dissipator and the algebra $\{L_\alpha\,,\,H\}'$ consisting of all matrices commuting with the larger 
set $\{L_\alpha\,,\,H\}$ obtained by adding to the the Lindblad operators also the system Hamiltonian, both coincide with the commutative algebra $\mathcal{M}=\{P,Q\}$ generated by the two orthogonal projections 
\begin{equation} 
P=\frac{1}{4} \left( \mathbb{I}_4 - \frac{S}{2} \right), \qquad Q=\mathbb{I}_4-P\ .
\end{equation}
It then follows that, given a faithful state $\hat\rho_0$ such that $\mathcal{L}[\hat\rho_0]=0$, the dynamics generated by $\mathcal{L}$ maps every given initial state $\rho(0)$ into an asymptotic, thus stationary, state of the form~\cite{frigerio}:
\begin{align}
\hat \rho = \frac{P \hat \rho_0 P}{\mathrm{Tr}[P \hat \rho_0]} \mathrm{Tr} [P \rho (0)] + \frac{Q \hat \rho_0 Q}{\mathrm{Tr}[Q \hat \rho_0]} \mathrm{Tr} [Q \rho (0)]. \label{family_asy}
\end{align}
\begin{rem}
\label{rem:Bellrep}
In terms of the Bell basis introduced in~\eqref{BellStates} of the main text,
the projector $P$ reads:
\begin{align}
 P= \frac{1}{4} \left( \mathbb{I}_4 - \frac{S}{2} \right)=|\psi_4 \rangle \langle \psi_4 |\ .
\end{align}
Therefore, with respect to the Bell basis, the two orthogonal projectors are represented by $P=\mathrm{diag}\{0,0,0,1\}$ and $Q=\mathrm{diag}\{1,1,1,0\}$.
Also, in the convex decomposition~\eqref{family_asy}, the coefficient reads
\begin{align}
\label{tau}
\mathrm{Tr} [P \rho (0)]= \frac{1-\tau}{4} \ ,
\end{align}
where $\rho (0)$ is the initial state and $\tau:= \sum_{i=1}^3 \rho_{ii}$ is the sum of the diagonal terms in the Fano decomposition (\ref{fano0}). 
\end{rem}
Substituting $\rho(0)=\hat\rho_0$ in \eqref{family_asy}, one finds 
$\hat \rho_0 = P \hat \rho_0 P + Q \hat \rho_0 Q$, so that the faithful state $\hat\rho_0$ decomposes into the orthogonal sum of two orthogonal matrices:
\begin{align}
\hat \rho_0 =\begin{pmatrix}
\rho_{11} & \rho_{12} & \rho_{13} & 0 \\
\rho_{12}^* & \rho_{22} & \rho_{23} & 0 \\
\rho_{13}^* & \rho_{23}^* & \rho_{33} & 0 \\
0&0&0&\rho_{44}
\end{pmatrix} \equiv \begin{pmatrix}
 &  & & 0 \\
& \,\,\hat\rho_0^Q &  & 0 \\
 &  &  & 0 \\
0&0&0&\rho_{44}
\end{pmatrix}.\label{asy_Bell}
\end{align}
As a consequence, the search for the faithful state $\hat \rho_0$ reduces to that of the component $\hat\rho^Q_0$, the enty  $\rho_{44}$ being then fixed by the normalization.
Furthermore, in the $3\times3$ subspace projected out by $Q$ and linearly spanned by the three Bell states $\lbrace \ket{\psi_1},\ket{\psi_2},\ket{\psi_3} \rbrace$, the operators $\lbrace\Sigma_i,\,S_{ij}\rbrace$ are represented by $\lbrace\Sigma_i^Q,\,S_{ij}^Q\rbrace$, where:
\begin{align}
\begin{split}
&\Sigma_1 ^Q =2\begin{pmatrix}
0 & 1 & 0 \\
1 & 0 & 0 \\
0 & 0 & 0
\end{pmatrix},\ \Sigma_2 ^Q =2\begin{pmatrix}
0 & 0 & 0 \\
0 & 0 & -i \\
0 & i & 0
\end{pmatrix}, \ \Sigma_3 ^Q =2\begin{pmatrix}
0 & 0 & 1 \\
0 & 0 & 0 \\
1 & 0 & 0
\end{pmatrix} \nonumber\\
&S_{11}^Q=2\begin{pmatrix}
1 & 0 & 0 \\
0 & 1 & 0 \\
0 & 0 & -1
\end{pmatrix}, \ S_{22}^Q=2\begin{pmatrix}
-1 & 0 & 0 \\
0 & 1 & 0 \\
0 & 0 & 1
\end{pmatrix}, \ S_{33}^Q=2\begin{pmatrix}
1 & 0 & 0 \\
0 & -1 & 0 \\
0 & 0 & 1
\end{pmatrix}\nonumber\\
&S_{12}^Q=2\begin{pmatrix}
0 & 0 & i \\
0 & 0 & 0 \\
-i & 0 & 0
\end{pmatrix},\ S_{13}^Q=2\begin{pmatrix}
0 & 0 & 0 \\
0 & 0 & 1 \\
0 & 1 & 0
\end{pmatrix}, \ S_{23}^Q=2\begin{pmatrix}
0 & i & 0 \\
-i & 0 & 0 \\
0 & 0 & 0
\end{pmatrix}.
\end{split}
\end{align}\\
It then follows that, while $\rho_{44}$ does not vary in time, the orthogonal component $\rho^Q$ obeys a master equation of the same form of \eqref{XdissME}, with the substitution $\{\Sigma_i,\,S_{ij}\}\rightarrow \{\Sigma_i^Q,\,S_{ij}^Q\}$:
\begin{align}\label{reduced_master_equation}
\nonumber\frac{\partial \rho^Q(t)}{\partial t}&=-i\left[\gamma \Sigma_{3}^Q+\sqrt{a}\,f\, S_{12}^Q, \rho^Q(t)\right]\\
&+\sum_{i, j=1}^{3}\left(\mathcal{A}_{i j}+\widetilde{\mathcal{A}}_{i j}\right)\left[\Sigma_{j}^Q \rho^Q(t) \Sigma_{i}^Q-\frac{1}{2}\left\{\Sigma_{i}^Q \Sigma_{j}^Q, \rho^Q(t)\right\}\right],
\end{align}
where the term $\delta S^Q$ does not appear in the Hamiltonian since $S^Q$ acts as the identity on the sub-space projected out by $Q$.
Such a projected master equation reduces the parameters needed to specify the faithful invariant state from 15, imposing unit trace and hermiticity, to 9 for $\hat\rho^Q_0$ since normalization is not fixed.

In the Bell-state representation, the entries of any faithful invariant 
$\hat\rho^Q_0$ in the kernel of the right hand side of~\eqref{reduced_master_equation}
can be analytically computed to be of the form 
\begin{align}
\hat \rho_0^Q = \begin{pmatrix}
A &0&W\\
0&B&0\\
W^*&0&C 
\end{pmatrix},
\end{align}
where $A,B,C \in \mathbb{R}$ and $W \in \mathbb{C}$, while the real part of $W$ is free:
a convenient $\hat\rho_0$ is selected by choosing $\rho_{44} =\rho_{22}$. Then, in the canonical basis, 
the faithful state becomes a particular $X$-state:
\begin{align}
\hat \rho_0=\begin{pmatrix}
a& 0&0&w\\
0&b&0&0\\
0&0&c&0\\
w^*&0&0&d\\
\end{pmatrix},
\end{align}\\
which, in terms of $\{ \mathbb{I}_4 , \Sigma_i , S_{ij}\}$, reads
\begin{equation} 
\label{faithful state Fano}
    \hat\rho_0 = \frac{1}{4} \bigg[\mathbb{I}_2 \otimes \mathbb{I}_2 + M \ \Sigma_3 -N \ (S_{11} - S_{22}) + R \ S_{33} - L \ S_{12} \bigg],
\end{equation}\\
where the coefficients are given by:
\begin{equation}
    M = -\frac{2 \sqrt{a} f \left(18 a^3+15 a^2 f^2+2 a \left(f^4+1\right)+f^2\right)}{36 a^4+60 a^3 f^2+a^2 \left(21 f^4+4\right)+2 a f^2 \left(f^4+2\right)+f^4}
\end{equation}
\begin{equation}
    N = \frac{a^2 f^2 \left(6 a+f^2\right)}{36 a^4+60 a^3 f^2+a^2 \left(21 f^4+4\right)+2 a f^2 \left(f^4+2\right)+f^4}
\end{equation}
\begin{equation}
    R = \frac{a f^2 \left(18 a^2+3 a f^2+2\right)}{36 a^4+60 a^3 f^2+a^2 \left(21 f^4+4\right)+2 a f^2 \left(f^4+2\right)+f^4}
\end{equation}
\begin{equation}
    L = \frac{4 a^2 f^2}{36 a^4+60 a^3 f^2+a^2 \left(21 f^4+4\right)+2 a f^2 \left(f^4+2\right)+f^4} \ .
\end{equation}
More in general, using~\eqref{family_asy} and setting ${\rm Tr}(P\,\rho(0))$ as in~\eqref{tau}, 
the convex asymptotic manifold consists of invariant states of the form 
\begin{align}
\hat \rho &=\frac{3+\tau}{4}\begin{pmatrix}
A&0&W&0\\
0&B&0&0\\
W&0&C&0\\
0&0&0&\displaystyle{\frac{1-\tau}{3+\tau}}
\end{pmatrix}\ , \label{faithful state lambda}
\end{align}
where 
\begin{align*}
 A&= \frac{\left(36 a^4+72 a^3 f^2+a^2 \left(23 f^4+4\right)+2 a f^2 \left(f^4+4\right)+f^4\right)}{108 a^4+216 a^3 f^2+3 a^2 \left(23 f^4+4\right)+2 a f^2 \left(3 f^4+8\right)+3 f^4}\ ,\\ 
 B&= \frac{\left(36 a^4+24 a^3 f^2+a^2 \left(15 f^4+4\right)+2 a f^6+f^4\right)}{108 a^4+216 a^3 f^2+3 a^2 \left(23 f^4+4\right)+2 a f^2 \left(3 f^4+8\right)+3 f^4}\ ,\\
 C&= \frac{ \left(36 a^4+120 a^3 f^2+a^2 \left(31 f^4+4\right)+2 a f^2 \left(f^4+4\right)+f^4\right)}{108 a^4+216 a^3 f^2+3 a^2 \left(23 f^4+4\right)+2 a f^2 \left(3 f^4+8\right)+3 f^4}\ ,\\
 W&= -\frac{4 \sqrt{a} f \left(2 i a^{3/2} f+18 a^3+15 a^2 f^2+2 a \left(f^4+1\right)+f^2\right)}{108 a^4+216 a^3 f^2+3 a^2 \left(23 f^4+4\right)+2 a f^2 \left(3 f^4+8\right)+3 f^4}\ .
\end{align*}
Any such invariant state is specified by the parameter $\tau$; for a given $\tau$, all initial two-qubit states with that $\tau$  are mapped to the corresponding invariant state which, in the standard representation, are generic $X$-states:
\begin{equation}
\label{Xaux}
\hat \rho=\begin{pmatrix}
a& 0&0&w\\
0&b&z&0\\
0&z^*&c&0\\
w^*&0&0&d
\end{pmatrix}\ .
\end{equation}

\section{Weak coupling limit constraint}
\label{AppB}

In this Appendix, we study the constraints on the unitary and dissipative parts of the Lindblad generator that are imposed by the weak coupling limit derivation of the master equation within the conditions specified in the main text.
These constraints amount to requesting that the unitary and dissipative terms commute and guarantee that the used generator can indeed be obtained by an actual microscopic coupling of the two qubits with a suitable environment.

The unitary term is generated by the Hamiltonian via
\begin{equation} \label{ham}
    \mathcal{H}[\rho]=-i[H,\rho],
\end{equation}
while the dissipative one acts through
\begin{equation} \label{dis}
    \mathcal{D}[\rho] = \sum_{i,j=1}^3 \mathcal{A}_{ij} \bigg[ \Sigma_j \rho \Sigma_i -\frac{1}{2} \left\{ \Sigma_i \Sigma_j , \rho \right\}\bigg],
\end{equation}
where $\mathcal{A}_{ij} = \mathrm{diag}\{a_{11},a_{22},a_{33}\}$ is the Kossakowski submatrix before feed-back control is performed.
Then, for the generator to be compatible with the weak-coupling limit, one must have that
\begin{equation}
\label{commaux0}
    \big[\mathcal{H},\mathcal{D} \big]=0 \ . 
\end{equation}
It is convenient to express the commutator with respect to the subspace generated by the first three Bell states $\{\ket{\psi_1}, \ket{\psi_2}, \ket{\psi_3}\}$ as done in Appendix~\ref{AppA}. Then, one writes the restriction
of a generic density matrix $\rho^Q(0)$ to that subspace as
\begin{align} \rho^Q(0)=
\left(
\begin{array}{ccc}
 \rho_{11} & \rho_{r12}+i\rho_{i12} & \rho_{r13}+i \rho_{i13} \\
 \rho_{r12}-i \rho_{i12} & \rho_{22} & \rho_{r23}+i \rho_{i23} \\
 \rho_{r13}-i\rho_{i13} & \rho_{r23}-i \rho_{i23} & \rho_{33} \\
\end{array}
\right)\ ,
\end{align}
where $\rho_{rab}$ and $\rho_{iab}$ stand for $\mathrm{Re}\{\rho_{ab}\}$ and $\mathrm{Im}\{\rho_{ab}\}$.
Represents it as a vector
$$
(\rho_{11},\rho_{22},\rho_{33},\rho_{r12},\rho_{i12},\rho_{r13},\rho_{i13},\rho_{r23},\rho_{i23})^T\in\mathbb{R}^9\ ,
$$
the action of the dissipative part of the generator~\eqref{dis} on the restricted state keeps it restricted to the subspace and can thus be represented by a $9\times9$ matrix 
\begin{align}
\label{matactaux1}
\mathcal{D}^Q_{9\times 9}=\begin{pmatrix}
 -4 (a_{11}+a_{33}) & 4 a_{11} & 4 a_{33} & 0 \\
 4 a_{11} & -4 (a_{11}+a_{22}) & 4 a_{22} & 0 \\
 4 a_{33} & 4 a_{22} & -4 (a_{22}+a_{33}) & 0 \\
 0 & 0 & 0 & R
\end{pmatrix}\ ,
\end{align}
where $R$ is the $6\times6$ diagonal matrix $R=-2\,\hbox{diag}\{r_i\}_{i=1}^6$, with
\begin{align}
\nonumber
&r_1=a_{22}+a_{33}\ ,\ r_2=4 a_{11}+a_{22}+a_{33}\ ,\ r_3=a_{11}+a_{22}\ ,\\ 
\label{matactaux2}
&r_4=a_{11}+a_{22}+4 a_{33}\ ,\ r_5=a_{11}+4a_{22}+a_{33}\ ,\ r_6=a_{11}+a_{33}\ .
\end{align}
On the other hand, given a $3\times 3$ Hamiltonian
\begin{align} 
\label{matactaux3}
H^Q =
    \left(
\begin{array}{ccc}
 h_{11} & h_{r12}+i h_{i12} & h_{r13}+i h_{i13} \\
 h_{r12}-i h_{i12} & h_{22} & h_{r23}+i h_{i23} \\
 h_{r13}-i h_{i13} & h_{r23}-i h_{i23} & h_{33} \\
\end{array}
\right)\ ,
\end{align}
the same procedure as before can be used to represent the action of~\eqref{ham} as the following $9\times 9$ 
matrix:
\begin{align}
\label{matactaux4}
&    \mathcal{H}_{9\times9} =
\begin{pmatrix}
 0 & A & B \\
 C & D & E\\
 F & G & H
\end{pmatrix}\ \hbox{where}\\
\nonumber
&A=\begin{pmatrix}
2 h_{i12} & -2 h_{r12} & 2 h_{i13}\\
-2 h_{i12} & 2 h_{r12} & 0\\
0 & 0 & -2 h_{i13}
\end{pmatrix}\ ,\ 
B=\begin{pmatrix}
-2 h_{r13} & 0 & 0\\
0 & 2 h_{i23} & -2 h_{r23}\\
2 h_{r13} & -2 h_{i23} & 2 h_{r23}
\end{pmatrix}\\
\nonumber
&C=\begin{pmatrix}
-h_{i12} & h_{i12} & 0\\
h_{r12} & -h_{r12} & 0\\
-h_{i13} & 0 & h_{i13}
\end{pmatrix}\ ,\
D=\begin{pmatrix}
0 & h_{11}-h_{22} & h_{i23}\\
h_{22}-h_{11} & 0 & h_{r23}\\
-h_{i23} & -h_{r23} & 0
\end{pmatrix}
\end{align}
and
\begin{align}
\nonumber
&E=\begin{pmatrix}
-h_{r23} & h_{i13} & -h_{r13}\\
h_{i23} & -h_{r13} & -h_{i13} \\
h_{11}-h_{33} & h_{i12} & h_{r12}
\end{pmatrix}\ ,\
F=\begin{pmatrix}
h_{r13} & 0 & -h_{r13}\\
0 & -h_{i23} &h_{i23}\\
0 & h_{r23} & -h_{r23}
\end{pmatrix}\ ,\\
\label{matactaux5}
&G=\begin{pmatrix}
h_{r23} & -h_{i23} & h_{33}-h_{11}\\
h_{r13} & -h_{i12} & h_{r12}\\
h_{i13} & -h_{r12} & -h{i12}
\end{pmatrix}\ ,\
H=\begin{pmatrix}
0 & -h_{r12} & h_{i12} \\
h_{r12} & 0 & h_{22}-h_{33}\\
-h{i12} & h_{33}-h_{22} & 0 
\end{pmatrix}\ .
\end{align}
Imposing $ \big[\mathcal{H}_{9\times9} \ , \ \mathcal{D}_{9\times9}\big]= 0$, one finds that the Hamiltonian compatible with the weak coupling limit must be of the form
\begin{equation} \label{reduced_ham_wcl}
    H^Q = \alpha \ \Sigma_1^Q + \beta \ \Sigma_2^Q + \gamma \ \Sigma_3^Q,
\end{equation}
with $\alpha, \beta, \gamma$ real parameters, and the Kossakowski matrix $\mathcal{A}$ must be a multiple of the identity, namely $a_{11}=a_{22}=a_{33}$.


\begin{thebibliography}{100}
\bibitem{breuer} H. P. Breuer, F. Petruccione, \emph{The theory of open quantum systems}, Oxford University Press (2002).
\bibitem{nielsen} M. A. Nielsen, I. L. Chuang, \emph{Quantum Computation and Quantum Information}, Massachusetts Institute of Technology (2010).
\bibitem{Alicki} R. Alicki, M. Fannes, \emph{Quantum Dynamical Systems}, Oxford University Press (2001).
\bibitem{benatti05} F. Benatti, R. Floreanini, \emph{Open quantum dynamics: complete positivity and entanglement}, International Journal of Modern Physics B \textbf{19}, 3063 (2005).
\bibitem{chrushinski17} D. Chrushinski, S. Pascazio, \emph{A Brief History of the GKLS Equation},  Open Systems \& Information Dynamics  \textbf{24}, 1740001 (2017).
\bibitem{gorini} V. Gorini, A. Kossakowski, and E. C. G. Sudarshan, \emph{Completely positive dynamical semi-groups of N -level systems}, J. Math. Phys. \textbf{17}, 821 (1976).
\bibitem{lindblad} G. Lindblad, \emph{On the generators of quantum dynamical semigroups}, Commun. Math. Phys. \textbf{48}, 119 (1976).

\bibitem{benatti04} F. Benatti, R. Floreanini, \emph{Entanglement generation in uniformly accelerating atoms: Reexamination of the Unruh effect}, Physical Review A \textbf{70}, 12112 (2004).
\bibitem{benatti05.1} F. Benatti, R. Floreanini, \emph{Controlling entanglement generation in
external quantum fields}, J. Opt. B: Quantum Semiclass. Opt. \textbf{7}, S429 (2005).
\bibitem{benatti06} F. Benatti, R. Floreanini, \emph{Asymptotic entanglement of two independent systems in a common bath}, International Journal of Quantum Information \textbf{4}, 395 (2006).
\bibitem{bouten} L. Bouten, R. Van Handel, M. R. James, \emph{An introduction to quantum filtering}, SIAM
Journal on Control and Optimization \textbf{46}, 2199 (2007).
\bibitem{gardiner_noise} C. W. Gardiner, P. Zoller, \emph{Quantum noise}, Springer (2000).
\bibitem{gough99} J. Gough, \emph{The Stratonovich Interpretation of Quantum Stochastic Approximations}, Potential Analysis \textbf{11}, 213 (1999).
\bibitem{wiseman10} H. M. Wiseman, G. J. Milburn, \emph{Quantum measurement and control}, Cambridge University Press (2010).
\bibitem{wiseman92} H. M. Wiseman and G. J. Milburn, \emph{Quantum theory and optical feed-back via homodyne detection}, Physical Review Letters \textbf{70}, 548 (1993).
\bibitem{wiseman93} H. M. Wiseman and G. J. Milburn, \emph{Quantum theory of field-quadrature measurements}, Physical Review A \textbf{47}, 642 (1993).
\bibitem{zhang17} J. Zhang, Y. Liu, R. Wu, K. Jacobs, F. Nori, \emph{Quantum feedback: theory, experiments and applications}, Physics Reports \textbf{679}, 1 (2017).
\bibitem{Spohn} H. Spohn, \emph{An algebraic condition for the approach to equilibrium of an open N-level system}, Lett. Math. Phys. \textbf{2}, 33 (1977).
\bibitem{parthasarathy92} K. R. Parthasarathy, \emph{An Introduction to Quantum Stochastic Calculus}, Birkhauser (1992).
\bibitem{wiseman02} H. M. Wiseman, S. Mancini, J. Wang, \emph{Bayesian feed-back versus Markovian feed-back in a two-level atom}, Physical Review A \textbf{66}, 13807 (2002).
\bibitem{wootters} W. K. Wootters, \emph{Entanglement of Formation of an Arbitrary State of Two Qubits}, Phys. Rev. Lett. \textbf{80}, 2245 (1998).
\bibitem{quesada} N. Quesada, A. Al-Qasimi, D. F. V. James. \emph{Quantum properties
and dynamics of X states},  Journal of Modern Optics \textbf{59}, 15 (2012).
\bibitem{frigerio} A. Frigerio, \emph{Stationary states of quantum dynamical semigroups}, Communications in Mathematical Physics \textbf{63}, 269–276 (1978).

\end{thebibliography}
\end{document}